# Rigorous Computation of Classical Horizontal Geodetic Networks

Sandi BERK[1] and Bojan STOPAR[2]

***Abstract***

*This paper examines mathematical models for processing classical horizontal geodetic (triangulation and trilateration) networks. Two rigorous parametric adjustment models are discussed. The first one is a well-known model of adjustment in the geodetic coordinate system. This model is completely rigorous (functional and stochastic parts) and uses unreduced distance and direction observations. The proposed alternative is a model of planar network adjustment with closed-form reductions of observations directly to the mapping plane. These ground-to-grid reductions are simple and universal, regardless of which map projection is used. Slightly different results of the planar network adjustment are obtained. The differences are attributed to a nonrigorous stochastic model. In theory, the stochastic properties of the reduced observations should also be adapted. However, these differences are very small and can always be neglected in geodetic and surveying practice.*

## Introduction

The concept and procedures of geodetic positioning have been fundamentally altered by the rise of global navigation satellite systems (GNSS) in the 1980s and 1990s. The focus of research has shifted from the classical concepts of geodetic positioning to concepts based on GNSS technology. Nevertheless, classical high-precision terrestrial geodetic networks can sometimes still be a good alternative, for example in ground deformation monitoring (e.g. Ruiz et al. 2003), or even indispensable in engineering surveying, especially in the case of underground control networks (e.g. Stengele and Schätti-Stählin 2010).

Classical horizontal and vertical geodetic networks, which are realized by terrestrial measurements, are traditionally separated and the so-called 2D plus 1D model is preferred to the 3D model, especially for high-precision surveys for engineering and geoscience projects (Kuang 1996, p. 42). Furthermore, in many areas – from scientific disciplines such as physics to practical applications in engineering surveys – it is acceptable practice to calculate in planar Cartesian coordinates (Chrisman 2017). Thus, the approach to the classical horizontal geodetic network computation in a projected coordinate system (henceforth: the conventional computational approach) is still widely used in geodetic and surveying practice. Its main disadvantage compared to the rigorous computation in the geodetic coordinate system is the inability to deal with networks spreading across two or more mapping zones (Shortis and Seager 1994).

Two rigorous parametric models for computation of coordinates of points in classical horizontal geodetic networks, also called triangulation/trilateration or triangulateration networks, are investigated in the present paper. A general remark on the use of the term 'rigorous' in this paper: It is assumed that a rigorous solution is given in a strict mathematical way by using closed-form equations, without approximations. However, iterative solutions are necessary for both models. Approaches to the computation of classical terrestrial geodetic networks from the seventies and eighties of the 20th century are examined, see e.g. Krakiwsky and Thomson (1978), Hradilek (1979), Vincenty (1980), and Mezera and Shrestha (1984). The solutions investigated in this paper are designed for computations:

[1] Surveying and Mapping Authority of the Republic of Slovenia, Zemljemerska ulica 12, SI-1000 Ljubljana, Slovenia ORCID: https://orcid.org/0000-0002-5074-6738 (corresponding author), email: sandi.berk@gov.si
[2] Prof., University of Ljubljana, Faculty of Civil and Geodetic Engineering, Jamova cesta 2, SI-1000 Ljubljana, Slovenia ORCID: https://orcid.org/0000-0003-3119-9967

- in the geodetic coordinate system (i.e., ellipsoidal adjustment model) and
- in projected coordinate systems, also known as the map grid or state plane coordinate systems (i.e., planar adjustment model).

Both solutions are based on the parametric model of three-dimensional geodetic network adjustment in the geodetic coordinate system. The main objective of this paper is a detailed presentation and performance analysis of a simple but rigorous functional model of horizontal geodetic network adjustment in the projected coordinate system. The model introduces strict ground-to-grid reductions of observations which are based on the corresponding observables. It was also implemented in the software (Berk 2008, pp. 22–26). Although the approach of direct reductions is known from literature (e.g. Vaníček and Krakiwsky 1986, p. 363), it seems to be somewhat overlooked. The conventional stepwise reductions are widely used in geodetic and surveying practice as well as in software solutions. After the submission of an early version of this paper in 2014, similar direct reductions were analyzed by Kadaj (2016). The same idea applied to the reduction of long spatial distances to a reference ellipsoid can be found in a proposal of Fotiou (1997). In the present paper, closed-form ground-to-grid reductions of distance and direction observations are combined with the height-controlled computational approach proposed by Vincenty (1980). A completely rigorous functional model of planar network adjustment is proposed – regardless of the map projection used. An insight is also given into the nonrigorous character of the stochastic model of the conventional planar network adjustment.

## Processing of Classical Horizontal Geodetic Networks

As already mentioned, classical terrestrial geodetic networks can be divided to horizontal and vertical ones. The vertical network must be solved first because the resulting heights of points are fixed in the computation of the horizontal geodetic network. To compute the latter rigorously, ellipsoidal heights of points are necessary (Sideris 1990); therefore, orthometric or normal heights must be transformed into ellipsoidal heights by using a geoid or quasi-geoid model. In the case of trigonometric heighting, the entire process – starting with the vertical network computation – should be repeated, since the accurate distances between the network points improve the accuracy of the calculated heights (Vincenty 1980).

The unknowns in the horizontal geodetic network adjustment are:

- coordinate unknowns – normally two for each new network point, but see also Vincenty's (1980) proposal of triplets of Cartesian coordinates $(X, Y, Z)$ and additional constraint equations, and
- orientation unknowns – normally one for each network point at which the directions are measured, but individual groups of sets of observations could also be processed separately by introducing more than one orientation unknown; if there is no common direction between them (e.g. in difficult weather conditions during measurement), an additional orientation unknown is indispensable.

The type of the coordinate unknowns depends on the coordinate system used for computation. Two types of coordinates are considered here:

- $(\lambda, \varphi)$ … geodetic longitudes and latitudes as coordinates of points in the geodetic coordinate system and
- $(e, n)$ … easting and northing coordinates of points in projected coordinate systems.

Three basic conventional terrestrial observables connecting the $i$th standpoint (an occupied network point) with the $j$th forepoint (a target point) – indexed by $i, j$ – are:

- $D_{i,j}$ ... spatial straight-line distance between the standpoint and the forepoint,
- $Z_{i,j}$ ... zenith distance at the standpoint, measured to the forepoint, and
- $A_{i,j}$ ... azimuth at the standpoint, measured to the forepoint.

The azimuth is usually not measured directly; when using the directional method (Ghilani and Wolf 2006, p. 101), the azimuth is expressed in terms of the horizontal direction observable (henceforth: direction observable) and the orientation unknown. The direction observable for the $k$th group of sets of observations at the standpoint is:

- $R_{i,j,k}$ ... horizontal direction at the standpoint, measured to the forepoint from the reference direction.

The input data for the horizontal geodetic network computation are distance and direction observations with the corresponding weight matrix.

It is assumed hereinafter that the raw observations are adequately preprocessed, resulting in:

- $D'_{i,j}$ ... mark-to-mark spatial straight-line distance observation and
- $R'_{i,j,k}$ ... direction observation, obtained as a reduced mean value of sets of horizontal directions in the local geodetic system.

The preprocessing of the original observations includes the transformation from the local astronomic to the local geodetic system due to the vertical deflections (Torge 2001, p. 243). This step is even more important after the introduction of the GRS80 ellipsoid associated with a new geocentric terrestrial geodetic datum (Featherstone and Rüeger 2000). The deflections of the vertical at the network points and the corresponding geoidal heights could be determined either by measurements (e.g. Hirt et al. 2010) or by modelling (e.g. Hirt 2010). In addition to the gravity field corrections mentioned above, all necessary meteorological corrections, instrumental calibration corrections and reductions to the mark-to-mark distances should also be carried out; see e.g. Kuang (1996, pp. 43–57). These preliminary reductions will not be discussed here to be able to focus on the main idea of the paper and present it as comprehensively as possible. However, the remarks at the end of this paper explain how the heights of instruments and targets as well as the vertical deflection components can also be rigorously introduced in the proposed adjustment model.

In order to determine a unique solution for the horizontal geodetic network computation based on redundant observations, least-squares adjustment technique should be applied. Assuming that the observations are normally distributed, this technique proves to provide the maximum likelihood estimation (e.g. Caspary 1988, p. 5). Using the notation of the well-known Gauss-Markov model, it is introduced as follows: $\mathrm{E}\{\boldsymbol{l}\} = \boldsymbol{A}\boldsymbol{x}$ and $\mathrm{D}\{\boldsymbol{l}\} = \sigma_0^2\boldsymbol{P}^{-1}$, where $\boldsymbol{l}$ is the vector of the observations, $\boldsymbol{x}$ is the vector of the network unknowns, $\boldsymbol{A}$ is the network configuration or design matrix, $\boldsymbol{P}$ is the weight matrix of observations, and $\sigma_0^2$ is the a priori variance factor. The estimated vector of the network unknowns is obtained as follows (Caspary 1988, p. 5):

$$\hat{\boldsymbol{x}} = \left(\boldsymbol{A}^{\mathrm{T}}\boldsymbol{P}\boldsymbol{A}\right)^{-1}\boldsymbol{A}^{\mathrm{T}}\boldsymbol{P}\boldsymbol{l} \tag{1}$$

For stochastically independent observations, the weight matrix $\boldsymbol{P}$ can be created as a diagonal matrix (Caspary 1988, p. 5). However, a fully populated inverse covariance matrix should be used for correlated observations. Stochastic independence in surveying practice is difficult to achieve. Correlations in a group of sets of horizontal directions, for example, have been investigated by Kregar et al. (2013).

Dealing with a linear least-squares problem generally requires linearization and an iterative solution technique (Teunissen 1990); a direct method for solving nonlinear problems is only possible in some special cases (e.g. Awange et al. 2003). Consequently, the vector of the observations $\boldsymbol{l}$ is calculated as the misclosure vector with the observed minus calculated values of observables. Similarly, the vector of the network unknowns $\boldsymbol{x}$ is calculated as the estimated minus approximate unknowns (Caspary 1988, p. 5).

## Rigorous Ellipsoidal Adjustment Model

One can start from the parametric model for the three-dimensional geodetic network computation in the geodetic coordinate system $(\lambda, \varphi, h)$ and, following the approach of Vincenty (1980), assume that the ellipsoidal heights of points $(h)$ are known values. Constants defining the size and shape of the reference ellipsoid are:

- $a$ … major semi-axis and
- $e$ … first numerical eccentricity.

Three parameters should be introduced first:

$$M_i = \frac{a(1-e^2)}{\sqrt{(1-e^2\sin^2\varphi_i)^3}} \tag{2}$$

$$N_i = \frac{a}{\sqrt{1-e^2\sin^2\varphi_i}} \tag{3}$$

$$\nu_i = \frac{dN_i}{d\varphi_i} = \frac{e^2 N_i \sin\varphi_i \cos\varphi_i}{1-e^2\sin^2\varphi_i} \tag{4}$$

$M_i$ is the radius of curvature of the meridian arc (meridian radius of curvature) and $N_i$ is the radius of curvature of the prime vertical (prime vertical radius of curvature) at the latitude $\varphi_i$, see e.g. Vaníček and Krakiwsky (1986, pp. 111–112 and p. 324) or Torge (2001, p. 96). $\nu_i$ is the derivative of the prime vertical radius of curvature with respect to the latitude $\varphi_i$.

Coordinate differences between the $i$th standpoint and the $j$th forepoint in the local geodetic system ($\Delta x_{i,j}$, $\Delta y_{i,j}$, and $\Delta z_{i,j}$) can be expressed in terms of the three basic conventional terrestrial observables ($D_{i,j}$, $Z_{i,j}$, and $A_{i,j}$) as follows (e.g. Hradilek 1979):

$$\Delta x_{i,j} = D_{i,j} \sin Z_{i,j} \cos A_{i,j} \tag{5}$$

$$\Delta y_{i,j} = D_{i,j} \sin Z_{i,j} \sin A_{i,j} \tag{6}$$

$$\Delta z_{i,j} = D_{i,j} \cos Z_{i,j} \tag{7}$$

It is appropriate to introduce some auxiliary parameters for the network baseline between the $i$th standpoint and the $j$th forepoint as follows:

$$\Delta\lambda_{i,j} = \lambda_j - \lambda_i \tag{8}$$

$$\alpha_{i,j} = \cos\varphi_i \sin\varphi_j - \sin\varphi_i \cos\varphi_j \cos\Delta\lambda_{i,j} \tag{9}$$

$$\beta_{i,j} = \sin\varphi_i \sin\varphi_j + \cos\varphi_i \cos\varphi_j \cos\Delta\lambda_{i,j} \tag{10}$$

The coordinate differences $\Delta x_{i,j}$, $\Delta y_{i,j}$, and $\Delta z_{i,j}$ in Eqs. (5) to (7) can now be expressed as follows:

$$\Delta x_{i,j} = \alpha_{i,j}\left(N_j + h_j\right) - e^2\left(N_j \sin\varphi_j - N_i \sin\varphi_i\right)\cos\varphi_i \tag{11}$$

$$\Delta y_{i,j} = \left(N_j + h_j\right)\cos\varphi_j \sin\Delta\lambda_{i,j} \tag{12}$$

$$\Delta z_{i,j} = \beta_{i,j}\left(N_j + h_j\right) - e^2\left(N_j \sin\varphi_j - N_i \sin\varphi_i\right)\sin\varphi_i - N_i - h_i \tag{13}$$

By solving the system of three Eqs. (5) to (7), one can obtain the observables expressed in terms of the coordinate unknowns $\lambda_i$, $\varphi_i$, $\lambda_j$, and $\varphi_j$ as follows:

$$D_{i,j} = \sqrt{\Delta x_{i,j}^2 + \Delta y_{i,j}^2 + \Delta z_{i,j}^2} \tag{14}$$

$$A_{i,j} = \mathrm{arctan2g}\left(\Delta y_{i,j}, \Delta x_{i,j}\right) \tag{15}$$

where the $\mathrm{arctan2g}$ function on the interval $[0, 2\pi)$ can – by adapting the example of Vermeille (2004) for the geodetic longitude determination – safely be calculated as

$$\mathrm{arctan2g}(y, x) = \begin{cases} \frac{\pi}{2} - 2\arctan\frac{x}{\sqrt{x^2+y^2}+y} & \ldots \quad y \geq 0 \\ \frac{3\pi}{2} + 2\arctan\frac{x}{\sqrt{x^2+y^2}-y} & \ldots \quad y < 0 \end{cases} \tag{16}$$

Eq. (16) is an improvement of the solution proposed by Meyer and Conshick (2014); the latter may lead to a division by zero for $\Delta y_{i,j} \cong 0 \ \land \ \Delta x_{i,j} \ll 0$.

As already mentioned, the corresponding direction observable in the local geodetic system is an indirect observable, which is defined as follows:

$$R_{i,j,k} = A_{i,j} - o_{i,k} \tag{17}$$

where $o_{i,k}$ is the orientation unknown for the respective ($k$th) group of direction observations at the $i$th standpoint; this is the angle between the geodetic north and the reference direction; see Fig. 1. Eqs. (14), (15), and (17) form the basis for a rigorous parametric model of adjustment of horizontal geodetic networks in the geodetic coordinate system.

Differences between the calculated values of observables $D_{i,j}$ and $R_{i,j,k}$ (based on the approximate values of the network unknowns) and the observations $D'_{i,j}$ and $R'_{i,j,k}$ – see the misclosure vector $\boldsymbol{l}$ in Eq. (1) – can be expressed as follows:

$$l_{D'_{i,j}} = D'_{i,j} - D_{i,j} = D'_{i,j} - \sqrt{\varDelta x_{i,j}^2 + \varDelta y_{i,j}^2 + \varDelta z_{i,j}^2} \tag{18}$$

$$l_{R'_{i,j,k}} = R'_{i,j,k} - R_{i,j,k} = R'_{i,j,k} - \operatorname{arctan2g}\left(\varDelta y_{i,j}, \varDelta x_{i,j}\right) + o_{i,k} \tag{19}$$

To avoid lengthy equations representing the elements of the network design matrix, these additional auxiliary parameters can be used for the network baseline connecting the $i$th standpoint and the $j$th forepoint:

$$\rho_{i,j} = \frac{\varDelta x_{i,j}}{\varDelta x_{i,j}^2 + \varDelta y_{i,j}^2} \quad \varsigma_{i,j} = \frac{\varDelta y_{i,j}}{\varDelta x_{i,j}^2 + \varDelta y_{i,j}^2} \qquad (20)–(21)$$

$$\chi_{i,j} = \frac{\varDelta x_{i,j}}{D_{i,j}} \quad \psi_{i,j} = \frac{\varDelta y_{i,j}}{D_{i,j}} \quad \xi_{i,j} = \frac{\varDelta z_{i,j}}{D_{i,j}} \qquad (22)–(24)$$

$$\varepsilon_{i,j} = \sin\varphi_i \cos\varphi_j - \cos\varphi_i \sin\varphi_j \cos\varDelta\lambda_{i,j} \tag{25}$$

$$\zeta_{i,j} = \cos\varphi_i \cos\varphi_j + \sin\varphi_i \sin\varphi_j \cos\varDelta\lambda_{i,j} \tag{26}$$

$$\eta_{i,j} = e^2\left(N_i \cos 2\varphi_i + \left(N_j \sin\varphi_j + \nu_i \cos\varphi_i\right)\sin\varphi_i\right) \tag{27}$$

$$\vartheta_{i,j} = \left(\nu_j \cos\varphi_j - \left(N_j + h_j\right)\sin\varphi_j\right)\sin\varDelta\lambda_{i,j} \tag{28}$$

$$\mu_{i,j} = N_j \cos\varphi_j + \nu_j \sin\varphi_j \tag{29}$$

$$\kappa_{i,j} = \zeta_{i,j}\left(N_j + h_j\right) + \nu_j \alpha_{i,j} - e^2 \mu_{i,j} \cos\varphi_i \tag{30}$$

Partial derivatives of the distance observable $D_{i,j}$ with respect to the network unknowns $\lambda_i$, $\varphi_i$, $\lambda_j$, and $\varphi_j$ can be expressed as follows:

$$\frac{\partial D_{i,j}}{\partial \lambda_i} = \left(\xi_{i,j}\cos\varphi_i - \chi_{i,j}\sin\varphi_i\right)\varDelta y_{i,j} - \psi_{i,j}\left(N_j + h_j\right)\cos\varphi_j \cos\varDelta\lambda_{i,j} \tag{31}$$

$$\frac{\partial D_{i,j}}{\partial \varphi_i} = \xi_{i,j}\varDelta x_{i,j} - \chi_{i,j}\left(\beta_{i,j}\left(N_j + h_j\right) - \eta_{i,j}\right) \tag{32}$$

$$\frac{\partial D_{i,j}}{\partial \lambda_j} = -\frac{\partial D_{i,j}}{\partial \lambda_i} \tag{33}$$

$$\frac{\partial D_{i,j}}{\partial \varphi_j} = \xi_{i,j}\left(\varepsilon_{i,j}\left(N_j + h_j\right) + \nu_j \beta_{i,j} - e^2 \mu_{i,j}\sin\varphi_i\right) + \psi_{i,j}\vartheta_{i,j} + \chi_{i,j}\kappa_{i,j} \tag{34}$$

Partial derivatives of the direction observable $R_{i,j,k}$ with respect to the network unknowns $\lambda_i$, $\varphi_i$, $\lambda_j$, $\varphi_j$, and $o_{i,k}$ can be expressed as follows:

$$\frac{\partial R_{i,j,k}}{\partial \lambda_i} = \varsigma_{i,j}\sin\varphi_i\,\varDelta y_{i,j} - \rho_{i,j}\left(N_j + h_j\right)\cos\varphi_j \cos\varDelta\lambda_{i,j} \tag{35}$$

$$\frac{\partial R_{i,j,k}}{\partial \varphi_i} = \varsigma_{i,j}\left(\beta_{i,j}\left(N_j + h_j\right) - \eta_{i,j}\right) \tag{36}$$

$$\frac{\partial R_{i,j,k}}{\partial \lambda_j} = -\frac{\partial R_{i,j,k}}{\partial \lambda_i} \tag{37}$$

$$\frac{\partial R_{i,j,k}}{\partial \varphi_j} = \rho_{i,j}\vartheta_{i,j} - \varsigma_{i,j}\kappa_{i,j} \tag{38}$$

$$\frac{\partial R_{i,j,k}}{\partial o_{i,k}} = -1 \tag{39}$$

All elements of the network design matrix for the network adjustment in the geodetic coordinate system – Eqs. (31) to (39) – are defined for any pair of intervisible points located on the Earth's surface and differing in horizontal coordinates.

The corresponding network design matrix $\boldsymbol{A_g}$ (subscript $\boldsymbol{g}$ is for geodetic coordinate system) can be created as follows:

$$\boldsymbol{A_g} = \begin{bmatrix} \ddots & \vdots & \vdots & \ddots & \vdots & \vdots & \ddots & \vdots & \vdots & \ddots & \vdots & \ddots \\ \cdots & \frac{\partial D_{i,j}}{\partial \lambda_i} & \frac{\partial D_{i,j}}{\partial \varphi_i} & \cdots & \frac{\partial D_{i,j}}{\partial \lambda_j} & \frac{\partial D_{i,j}}{\partial \varphi_j} & \cdots & 0 & 0 & \cdots & 0 & \cdots \\ \ddots & \vdots & \vdots & \ddots & \vdots & \vdots & \ddots & \vdots & \vdots & \ddots & \vdots & \ddots \\ \cdots & 0 & 0 & \cdots & \frac{\partial R_{j,k,l}}{\partial \lambda_j} & \frac{\partial R_{j,k,l}}{\partial \varphi_j} & \cdots & \frac{\partial R_{j,k,l}}{\partial \lambda_k} & \frac{\partial R_{j,k,l}}{\partial \varphi_k} & \cdots & \frac{\partial R_{j,k,l}}{\partial o_{i,l}} & \cdots \\ \ddots & \vdots & \vdots & \ddots & \vdots & \vdots & \ddots & \vdots & \vdots & \ddots & \vdots & \ddots \end{bmatrix} \tag{40}$$

where $D_{i,j}$ refers to the spatial straight-line distance between the $i$th and $j$th network point and $R_{j,k,l}$ refers to the direction observable from the $j$th standpoint to the $k$th forepoint, obtained within the $l$th group of observations. The vector of the network unknowns $\boldsymbol{\hat{x}_g}$ is estimated using Eq. (1).

For the test network presented in the second part of this article it is assumed that the a priori variance factor is not reliably known. Therefore, the a posteriori variance factor $\hat{\sigma}_0^2$ is used here instead of the a priori variance factor $\sigma_0^2$, as usual (e.g. Kuang 1996, p. 165; Caspary 1988, p 40), to estimate the accuracy of the network unknowns. The covariance matrix of the estimated network unknowns from the adjustment in the geodetic coordinate system is therefore expressed as:

$$\boldsymbol{\Sigma}_{\boldsymbol{\hat{x}_g}} = \hat{\sigma}_0^2 \left(\boldsymbol{A_g}^{\mathrm{T}} \boldsymbol{P} \boldsymbol{A_g}\right)^{-1} \tag{41}$$

with the estimated a posteriori variance factor defined as:

$$\hat{\sigma}_0^2 = \frac{\boldsymbol{v}^T \boldsymbol{P} \boldsymbol{v}}{r} \tag{42}$$

where $r$ is the number of redundant observations and $\boldsymbol{v}$ is the residual vector obtained as follows:

$$\boldsymbol{v} = \boldsymbol{A_g} \boldsymbol{\hat{x}_g} - \boldsymbol{l} \tag{43}$$

The resulting accuracy estimates of the network points refer to the orthogonal curvilinear geodetic coordinates $(\lambda, \varphi)$. For the transformation on a differential manifold between curvilinear and linear geodetic coordinates $(\bar{e}, \bar{n})$ the so-called metric or Lamé matrix $\boldsymbol{H}_i$ is used as follows:

$$\begin{bmatrix} \bar{e}_i \\ \bar{n}_i \end{bmatrix} = \boldsymbol{H}_i \begin{bmatrix} \lambda_i \\ \varphi_i \end{bmatrix} \tag{44}$$

which is modified for the horizontal geodetic network by omitting the height component (cf. Soler and Smith 2010):

$$\boldsymbol{H}_i = \begin{bmatrix} N_i \cos \varphi_i & 0 \\ 0 & M_i \end{bmatrix} \tag{45}$$

where $M_i$ and $N_i$ are the principal radii of curvature from Eqs. (2) and (3). In this way, the accuracy estimates obtained are measured in linear units and refer to the footpoint on the reference ellipsoid. The corresponding block matrix $\boldsymbol{H}$, which is here referred to as the network metric matrix, can be created as follows:

$$
\boldsymbol{H} = \begin{bmatrix} [\boldsymbol{H}_1] & & \dots & \begin{matrix} 0 & 0 \\ 0 & 0 \end{matrix} & \dots & \begin{matrix} 0 \\ 0 \end{matrix} & \dots & \begin{matrix} 0 \\ 0 \end{matrix} \\ \vdots & & \ddots & \vdots & \ddots & \vdots & \ddots & \vdots \\ \begin{matrix} 0 & 0 \\ 0 & 0 \end{matrix} & & \dots & [\boldsymbol{H}_i] & \dots & \begin{matrix} 0 \\ 0 \end{matrix} & \dots & \begin{matrix} 0 \\ 0 \end{matrix} \\ \vdots & & \ddots & \vdots & \ddots & \vdots & \ddots & \vdots \\ 0 & 0 & \dots & 0 \quad 0 & \dots & 1 & \dots & 0 \\ \vdots & & \ddots & \vdots & \ddots & \vdots & \ddots & \vdots \\ 0 & 0 & \dots & 0 \quad 0 & \dots & 0 & \dots & 1 \end{bmatrix} \tag{46}
$$

where the lower right identity sub-matrix is of the size corresponding to the number of orientation unknowns, and the diagonal sub-matrix $\boldsymbol{H}_i$ is the Lamé matrix for the $i$th new network point, Eq. (45). The corresponding covariance matrix (in linear units for the coordinate unknown estimates) is:

$$
\boldsymbol{\Sigma}_{\hat{\boldsymbol{x}}_{lg}} = \hat{\sigma}_0^2 \boldsymbol{H} \left( \boldsymbol{A}_{\boldsymbol{g}}^{\mathrm{T}} \boldsymbol{P} \boldsymbol{A}_{\boldsymbol{g}} \right)^{-1} \boldsymbol{H}^{\mathrm{T}} \tag{47}
$$

and can be written as:

$$
\boldsymbol{\Sigma}_{\hat{\boldsymbol{x}}_{lg}} = \begin{bmatrix} \ddots & & \vdots & \ddots & \vdots & \ddots \\ \dots & & [\boldsymbol{\Sigma}_i] & \dots & \begin{matrix} \hat{\sigma}_{\bar{e}_i o_{j,l}} \\ \hat{\sigma}_{\bar{n}_i o_{j,l}} \end{matrix} & \dots \\ \ddots & & \vdots & \ddots & \vdots & \ddots \\ \dots & \hat{\sigma}_{\bar{e}_i o_{j,l}} & \hat{\sigma}_{\bar{n}_i o_{j,l}} & \dots & \hat{\sigma}_{o_{j,l}}^2 & \dots \\ \ddots & & \vdots & \ddots & \vdots & \ddots \end{bmatrix} \tag{48}
$$

where the lower right part of the diagonal contains the estimated variances of the orientation unknowns, e.g. for the $j$th standpoint and $l$th group of observations, and the $i$th diagonal sub-matrix $\boldsymbol{\Sigma}_i$ contains the estimated variances and covariances of the pair of coordinates for the $i$th new network point:

$$
\boldsymbol{\Sigma}_i = \begin{bmatrix} \hat{\sigma}_{\bar{e}_i}^2 & \hat{\sigma}_{\bar{e}_i \bar{n}_i} \\ \hat{\sigma}_{\bar{e}_i \bar{n}_i} & \hat{\sigma}_{\bar{n}_i}^2 \end{bmatrix} \tag{49}
$$

The above matrix elements refer to the local geodetic system and are given in linear units. The corresponding standard confidence ellipse elements for the $i$th network point are (e.g. Kuang 1996, pp. 164–168):

$$
\bar{q}_i = \sqrt{\left( \hat{\sigma}_{\bar{n}_i}^2 - \hat{\sigma}_{\bar{e}_i}^2 \right)^2 + 4 \hat{\sigma}_{\bar{e}_i \bar{n}_i}^2} \tag{50}
$$

$$
\bar{a}_i = \frac{1}{\sqrt{2}} \sqrt{\hat{\sigma}_{\bar{e}_i}^2 + \hat{\sigma}_{\bar{n}_i}^2 + \bar{q}_i} \quad \bar{b}_i = \frac{1}{\sqrt{2}} \sqrt{\hat{\sigma}_{\bar{e}_i}^2 + \hat{\sigma}_{\bar{n}_i}^2 - \bar{q}_i} \tag{51–52}
$$

$$
\bar{t}_i = \begin{cases} 0 & \dots \ \hat{\sigma}_{\bar{e}_i}^2 = \hat{\sigma}_{\bar{n}_i}^2 \ \land \ \hat{\sigma}_{\bar{e}_i \bar{n}_i} = 0 \\ \frac{1}{2} \operatorname{arctan2g}\left( 2 \hat{\sigma}_{\bar{e}_i \bar{n}_i}, \hat{\sigma}_{\bar{n}_i}^2 - \hat{\sigma}_{\bar{e}_i}^2 \right) & \dots \ \hat{\sigma}_{\bar{e}_i}^2 \neq \hat{\sigma}_{\bar{n}_i}^2 \ \lor \ \hat{\sigma}_{\bar{e}_i \bar{n}_i} \neq 0 \end{cases} \tag{53}
$$

where $\operatorname{arctan2g}$ is defined by Eq. (16), $\bar{q}_i$ is an auxiliary parameter, $\bar{a}_i$ is the major semi-axis, $\bar{b}_i$ is the minor semi-axis, and $\bar{t}_i$ is the azimuth of the major semi-axis, which is given here on the interval $[0, \pi)$.

## Rigorous Planar Adjustment Model

The conventional computational approach to determine the coordinates of points of a horizontal geodetic network in a projected coordinate system $(e, n)$ involves (Torge 2001, p. 311):

- the reduction of terrestrial observations into the projected coordinate system and
- the adjustment of reduced observations as if they were measured on a flat Earth.

The conventional computational approach generally requires a conformal mapping (e.g. Kuang 1996, p. 59). Starting with the mark-to-mark corrected observations in the local geodetic system, three additional steps of geometric reduction of observations are required (e.g. Vaníček and Krakiwsky 1986, pp. 348–352 and 361–362; Kuang 1996, pp. 56–62; Torge 2001, pp. 243–245).

The first step comprises the reduction of observations to the reference ellipsoid. The distance observation is first reduced from the spatial straight line between the standpoint and the forepoint to the normal section on the reference ellipsoid – its intersection with the plane containing the normal through the standpoint and the footpoint of the forepoint (spatial straight line to normal section distance reduction). The direction observation is first reduced from the horizontal direction in the local geodetic system pointing to the forepoint to the direction pointing to the footpoint of the forepoint on the reference ellipsoid (skew-normal direction reduction).

The second and third steps of geometric reduction of observations are:

- reductions of observations from the normal section on the reference ellipsoid to the geodesic between the footpoint of the standpoint and the footpoint of the forepoint (normal section to geodesic reductions) and
- reductions of observations from the surface of the reference ellipsoid to the mapping plane; simply explained as the reductions from the geodesic to the straight line between the projected standpoint and the projected forepoint on the mapping plane (arc-to-chord reductions).

The change of the coordinate system for the coordinate unknowns can be realized by applying the corresponding mapping equations as follows:

$$e_i = \mathrm{e}(\lambda_i, \varphi_i) \tag{54}$$

$$n_i = \mathrm{n}(\lambda_i, \varphi_i) \tag{55}$$

The inverse mapping equations needed to calculate the geodetic coordinates of a point from its projected coordinates can be written formally as:

$$\lambda_i = \mathrm{\lambda}(e_i, n_i) \tag{56}$$

$$\varphi_i = \mathrm{\varphi}(e_i, n_i) \tag{57}$$

Let the coordinate differences of adjacent network points be denoted as:

$$\varDelta e_{i,j} = e_j - e_i \tag{58}$$

$$\varDelta n_{i,j} = n_j - n_i \tag{59}$$

The observables in the projected coordinate system are grid distance $\widetilde{D}_{i,j}$ and grid azimuth $\tilde{A}_{i,j}$ of the chord (e.g. Vaníček and Krakiwsky 1986, pp. 403–404; Ghilani and Wolf 2006, p. 236 and 256):

$$\widetilde{D}_{i,j} = \sqrt{\varDelta e_{i,j}^2 + \varDelta n_{i,j}^2} \tag{60}$$

$$\tilde{A}_{i,j} = \mathrm{arctan2g}\left(\varDelta e_{i,j}, \varDelta n_{i,j}\right) \tag{61}$$

with arctan2g defined by Eq. (16). Grid azimuth or grid bearing $\tilde{A}_{i,j}$ is an indirect observable. However, instead of introducing the conventional orientation unknown, see $o_{i,k}^c$ in Fig. 1, the relationship between the grid azimuth and the selected direction observable $\tilde{R}_{i,j,k}$ can be defined as:

$$\tilde{R}_{i,j,k} = \tilde{A}_{i,j} - o_{i,k} \tag{62}$$

where $o_{i,k}$ is the orientation unknown for the respective ($k$th) group of direction observations at the $i$th standpoint – the same orientation unknown is used for the model of the network adjustment in the geodetic coordinate system, which is presented in the previous section. This is a crucial point of the approach proposed in this paper.

Fig. 1 explains relations between the direction observables in both network adjustment models. The left side shows the approach used for the orientation of the horizontal directions in the local geodetic system: the direction observable ($R_{i,j,k}$) and the geodetic azimuth ($A_{i,j}$) are connected by the orientation unknown ($o_{i,k}$) – see Eq. (17), where:

- $\mathrm{P}_i$ and $\mathrm{P}_j$ … are the $i$th standpoint and the $j$th forepoint,
- N … indicates the direction of the meridian through the standpoint due true or geodetic north ($x$ axis of the local geodetic system),
- O … indicates the reference direction of the horizontal circle of the theodolite,
- F … indicates the (measured) direction to the forepoint, and
- F' … indicates the direction of the tangent to the geodesic towards the footpoint of the forepoint on the reference ellipsoid.

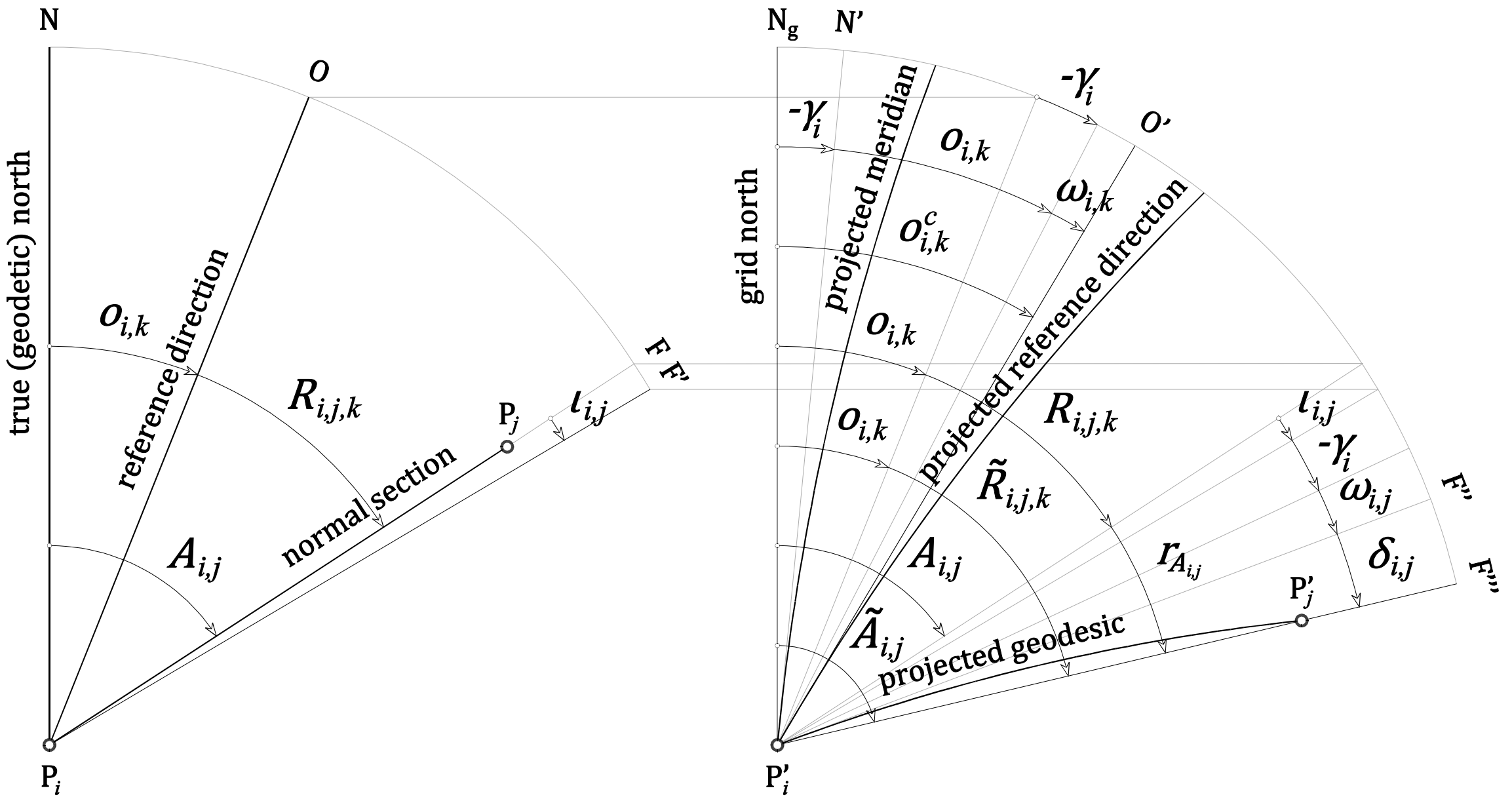

***Fig. 1.*** *Relations between the direction observables in geodetic (left) and projected coordinate systems (right)*

The right side of Fig. 1 shows the approach used for the orientation of reduced horizontal directions in a projected coordinate system (not necessarily on a conformal mapping plane): the selected direction observable ($\tilde{R}_{i,j,k}$) and grid azimuth ($\tilde{A}_{i,j}$) are connected by the same orientation unknown ($o_{i,k}$) – see Eq. (62), where:

- $\mathrm{P}'_i$ and $\mathrm{P}'_j$ … are the projected $i$th standpoint and the projected $j$th forepoint,
- $\mathrm{N_g}$ … indicates the direction, which is parallel to the ordinate axis ($n$ axis) of the projected coordinate system through the standpoint,
- N' … indicates the direction of the tangent to the projected meridian through the standpoint due north,
- O' … indicates the direction of the tangent to the projected reference direction of the theodolite, and
- F'' … indicates the direction of the tangent to the projected geodesic towards the footpoint of the forepoint on the reference ellipsoid, and
- F''' … indicates the direction of the chord of the projected geodesic towards the footpoint of the forepoint on the reference ellipsoid.

The remaining quantities in Fig. 1 are:

- $\gamma_i$ … meridian or grid convergence at the $i$th standpoint, measured from the true north (e.g. Vaníček and Krakiwsky 1986, p. 361) – therefore the minus sign in Fig. 1,
- $\omega_{i,j}$ … angular distortion at the $i$th standpoint for the direction to the $j$th forepoint, which is zero in the case of conformal mapping,

- $\omega_{i,k}$ … angular distortion at the $i$th standpoint for the reference direction in the $k$th group of observations, which is zero in the case of conformal mapping,
- $\iota_{i,j}$ … skew-normal reduction plus normal section to geodesic reduction at the $i$th standpoint for the direction to the $j$th forepoint (e.g. Torge 2001, pp. 243–244),
- $\delta_{i,j}$ … arc-to-chord reduction at the $i$th standpoint for the direction to the $j$th forepoint (e.g. Kuang 1996, p. 61),
- $o_{i,k}^{c} = o_{i,k} - \gamma_i + \omega_{i,k}$ … orientation unknown as defined in the conventional computational approach (with $\omega_{i,k} = 0$ as a rule), and
- $r_{A_{i,j}} = \tilde{A}_{i,j} - A_{i,j} = \iota_{i,j} - \gamma_i + \omega_{i,j} + \delta_{i,j}$ … the total one-step reduction at the $i$th standpoint for the direction to the $j$th forepoint, as proposed in this paper.

The signs of individual reductions may also vary according to their definition by different authors. The selected orientation unknown ($o_{i,k}$) does not match the corresponding orientation unknown in the conventional computational approach ($o_{i,k}^{c}$) – unless a conformal ($\omega_{i,k} = 0$) cylindrical ($\gamma_i = 0$) projection is used (e.g. normal Mercator projection). In the case of nonconformal mapping, the formula for the maximum angular distortion at the $i$th point ($\omega_i$) can be derived (e.g. Snyder 1987, pp. 20–24); however, the azimuth-dependent angular distortion ($\omega_{i,j}$) is not easy to handle.

Details of geometric reduction of the distance observations will not be discussed here. An advantage of the proposed definitions of orientation angles and direction observables is that reductions can be rigorously applied to both types of observables. By applying Eqs. (14), (15), (17), and (60) to (62) – see also Fig. 1 – these reductions can be expressed as:

$$r_{D_{i,j}} = \widetilde{D}_{i,j} - D_{i,j} = \sqrt{\Delta e_{i,j}^2 + \Delta n_{i,j}^2} - \sqrt{\Delta x_{i,j}^2 + \Delta y_{i,j}^2 + \Delta z_{i,j}^2} \tag{63}$$

$$r_{A_{i,j}} = \tilde{A}_{i,j} - A_{i,j} = \tilde{R}_{i,j,k} - R_{i,j,k} = \mathrm{arctan2g}(\Delta e_{i,j}, \Delta n_{i,j}) - \mathrm{arctan2g}(\Delta y_{i,j}, \Delta x_{i,j}) \tag{64}$$

The above reductions (based on the corresponding observables) are close but not equal to the conventional stepwise reductions of observations. Also, the latter involve a certain degree of approximation. For example, even the most accurate straightforward reduction of the spatial distances to a reference ellipsoid can lead to errors reaching up to ~0.1 mm (Thomson and Vaníček 1974). The formulas for the traditional arc-to-chord distance and direction reductions also depend on the mapping equations and involve working with the parametric equations of the projected geodesic and evaluating its curvature (Vaníček and Krakiwsky 1986, pp. 362–363). The advantage of the proposed Eqs. (63) and (64) is therefore that they are simple, strict (i.e., closed-form equations are applied), and universal, regardless of which map projection is used – even nonconformal mapping is allowed. Reductions of observations from the local geodetic system directly into the projected coordinate system can now be performed as follows:

$$\widetilde{D}'_{i,j} = D'_{i,j} + r_{D_{i,j}} \tag{65}$$

$$\tilde{R}'_{i,j,k} = R'_{i,j,k} + r_{A_{i,j}} \tag{66}$$

The differences between the calculated values of observables and the respective reduced distance and direction observations can be expressed as follows:

$$l_{\widetilde{D}'_{i,j}} = \widetilde{D}'_{i,j} - \widetilde{D}_{i,j} = \widetilde{D}'_{i,j} - \sqrt{\Delta e_{i,j}^2 + \Delta n_{i,j}^2} \tag{67}$$

$$l_{\tilde{R}'_{i,j,k}} = \tilde{R}'_{i,j,k} - \tilde{R}_{i,j,k} = \tilde{R}'_{i,j,k} - \mathrm{arctan2g}(\Delta e_{i,j}, \Delta n_{i,j}) + o_{i,k} \tag{68}$$

By inserting Eqs. (65) and (66) for both reductions, these misclosures can be expressed in a way that has already been seen in the previous section – compare with Eqs. (18) and (19):

$$l_{\widetilde{D}'_{i,j}} = D'_{i,j} - \sqrt{\Delta x_{i,j}^2 + \Delta y_{i,j}^2 + \Delta z_{i,j}^2} \tag{69}$$

$$l_{\tilde{R}'_{i,j,k}} = R'_{i,j,k} - \mathrm{arctan2g}(\Delta y_{i,j}, \Delta x_{i,j}) + o_{i,k} \tag{70}$$

The only additional step required to apply Eqs. (69) and (70) is the simultaneous conversion of the improved projected coordinates into the improved geodetic coordinates in each iteration step using the corresponding inverse mapping Eqs. (56) and (57).

At this point, it can be emphasized that any definition of the orientation angle other than the proposed $o_{i,k}$ would have some disadvantages for further theoretical considerations – see the mapping design matrix, Eq. (83). In the case of conformal mapping ($\omega_{i,k} = 0$) and the conventional orientation angle ($o_{i,k}^c = o_{i,k} - \gamma_i$), see Fig. 1, the mapping design matrix would obtain additional nonzero off-diagonal elements with partial derivatives of the meridian convergence from Eq. (86) with respect to the geodetic longitude and latitude. These elements contain the second-order partial and mixed derivatives of the mapping functions from Eqs. (54) and (55). The presented model is simpler and more general, since it requires only the first-order derivatives of the projected coordinates with respect to the geodetic coordinates; in case of nonconformal mapping, the matter would be even more complicated.

The elements of the network design matrix $\boldsymbol{A}$ for the planar adjustment shall be repeated here; see e.g. Mikhail and Gracie (1981, pp. 266–272) or Ghilani and Wolf (2006, p. 237 and p. 257). The partial derivatives of the reduced distance observable $\widetilde{D}_{i,j}$ with respect to the network unknowns $e_i$, $n_i$, $e_j$, and $n_j$ can be expressed as follows:

$$\frac{\partial \widetilde{D}_{i,j}}{\partial e_i} = \frac{-\Delta e_{i,j}}{\widetilde{D}_{i,j}} \quad \frac{\partial \widetilde{D}_{i,j}}{\partial n_i} = \frac{-\Delta n_{i,j}}{\widetilde{D}_{i,j}} \tag{71)–(72}$$

$$\frac{\partial \widetilde{D}_{i,j}}{\partial e_j} = -\frac{\partial \widetilde{D}_{i,j}}{\partial e_i} \quad \frac{\partial \widetilde{D}_{i,j}}{\partial n_j} = -\frac{\partial \widetilde{D}_{i,j}}{\partial n_i} \tag{73)–(74}$$

Partial derivatives of the reduced direction observable $\tilde{R}_{i,j,k}$ with respect to the network unknowns $e_i$, $n_i$, $e_j$, $n_j$, and $o_{i,k}$ can be expressed as follows:

$$\frac{\partial \tilde{R}_{i,j,k}}{\partial e_i} = \frac{-\Delta n_{i,j}}{\widetilde{D}_{i,j}^2} \quad \frac{\partial \tilde{R}_{i,j,k}}{\partial n_i} = \frac{\Delta e_{i,j}}{\widetilde{D}_{i,j}^2} \tag{75)–(76}$$

$$\frac{\partial \tilde{R}_{i,j,k}}{\partial e_j} = -\frac{\partial \tilde{R}_{i,j,k}}{\partial e_i} \quad \frac{\partial \tilde{R}_{i,j,k}}{\partial n_j} = -\frac{\partial \tilde{R}_{i,j,k}}{\partial n_i} \quad \frac{\partial \tilde{R}_{i,j,k}}{\partial o_{i,k}} = -1 \tag{77)–(79}$$

The corresponding network design matrix $\boldsymbol{A_p}$ (subscript $\boldsymbol{p}$ is for projected coordinate system) can be created as follows:

$$\boldsymbol{A_p} = \begin{bmatrix} \ddots & \vdots & \vdots & \ddots & \vdots & \vdots & \ddots & \vdots & \vdots & \ddots & \vdots & \ddots \\ \cdots & \frac{\partial \widetilde{D}_{i,j}}{\partial e_i} & \frac{\partial \widetilde{D}_{i,j}}{\partial n_i} & \cdots & \frac{\partial \widetilde{D}_{i,j}}{\partial e_j} & \frac{\partial \widetilde{D}_{i,j}}{\partial n_j} & \cdots & 0 & 0 & \cdots & 0 & \cdots \\ \ddots & \vdots & \vdots & \ddots & \vdots & \vdots & \ddots & \vdots & \vdots & \ddots & \vdots & \ddots \\ \cdots & 0 & 0 & \cdots & \frac{\partial \tilde{R}_{j,k,l}}{\partial e_j} & \frac{\partial \tilde{R}_{j,k,l}}{\partial n_j} & \cdots & \frac{\partial \tilde{R}_{j,k,l}}{\partial e_k} & \frac{\partial \tilde{R}_{j,k,l}}{\partial n_k} & \cdots & \frac{\partial \tilde{R}_{j,k,l}}{\partial o_{i,l}} & \cdots \\ \ddots & \vdots & \vdots & \ddots & \vdots & \vdots & \ddots & \vdots & \vdots & \ddots & \vdots & \ddots \end{bmatrix} \tag{80}$$

where $\widetilde{D}_{i,j}$ refers to the chord distance between the $i$th and $j$th network point and $\tilde{R}_{j,k,l}$ refers to the direction observable from the $j$th standpoint to the $k$th forepoint, obtained within the $l$th group of observations. In this way, a rigorous functional model of adjustment in the projected coordinate system is created. The vector of the network unknowns $\widehat{\widehat{\boldsymbol{x}}}_{\boldsymbol{p}}$ is estimated with Eq. (1) by introducing the network design matrix from Eq. (80).

Again, it is assumed that the a priori variance factor is not reliably known – see remarks before Eq. (41) –, therefore the covariance matrix of the estimated network unknowns from the adjustment in the projected coordinate system is expressed as:

$$\boldsymbol{\Sigma}_{\widehat{\widehat{x}}_{\boldsymbol{p}}} = \widehat{\widehat{\sigma}}_0^2 \left(\boldsymbol{A_p}^{\mathrm{T}} \boldsymbol{P} \boldsymbol{A_p}\right)^{-1} \tag{81}$$

with the a posteriori variance factor $\hat{\hat{\sigma}}_0^2$ determined in the same way as in the previous section, see Eq. (42), but using observation residuals from the planar network adjustment. It should be noted that the observation residuals from this planar network adjustment differ slightly from those obtained with Eq. (43), which is due to the reduction of observations to the mapping plane.

The standard confidence ellipses of the new network points in the projected coordinate system can now be determined by using the covariance matrix elements from Eq. (81) and applying Eqs. (50) to (53).

### *Conversion of the Computation Results into a Projected Coordinate System*

In order to be able to compare the results of the presented planar model of the computation of horizontal geodetic networks ($\hat{\hat{\boldsymbol{x}}}_{\boldsymbol{p}}$, $\boldsymbol{\Sigma}_{\hat{\hat{x}}_p}$) with the results of the computation in the geodetic coordinate system, the latter should be rigorously converted into a projected coordinate system ($\hat{\boldsymbol{x}}_{\boldsymbol{p}}$, $\boldsymbol{\Sigma}_{\hat{x}_p}$). The estimated coordinate unknowns ($\hat{\boldsymbol{x}}_{\boldsymbol{g}}$) can be converted with the corresponding mapping Eqs. (54) and (55). To adapt the corresponding covariance matrix of the estimated network unknowns ($\boldsymbol{\Sigma}_{\hat{x}_g}$) obtained with Eq. (41), the law of variance-covariance propagation (e.g. Mikhail and Gracie 1981, pp. 152–154) can be used as follows:

$$\boldsymbol{\Sigma}_{\hat{x}_p} = \boldsymbol{A}_{\boldsymbol{m}} \boldsymbol{\Sigma}_{\hat{x}_g} \boldsymbol{A}_{\boldsymbol{m}}^{\mathrm{T}} \tag{82}$$

where $\boldsymbol{A}_{\boldsymbol{m}}$ is referred to as the mapping design matrix and can be created as follows:

$$\boldsymbol{A}_{\boldsymbol{m}} = \begin{bmatrix} \multicolumn{2}{c}{[J_1]} & \cdots & 0 & 0 & \cdots & 0 & \cdots & 0 \\ & & & 0 & 0 & & 0 & & 0 \\ \vdots & & \ddots & \vdots & & \ddots & \vdots & \ddots & \vdots \\ 0 & 0 & \cdots & \multicolumn{2}{c}{[J_i]} & \cdots & 0 & \cdots & 0 \\ 0 & 0 & & & & & 0 & & 0 \\ \vdots & & \ddots & \vdots & & \ddots & \vdots & \ddots & \vdots \\ 0 & 0 & \cdots & 0 & 0 & \cdots & 1 & \cdots & 0 \\ \vdots & & \ddots & \vdots & & \ddots & \vdots & \ddots & \vdots \\ 0 & 0 & \cdots & 0 & 0 & \cdots & 0 & \cdots & 1 \end{bmatrix} \tag{83}$$

and where the lower right identity sub-matrix is of the size corresponding to the number of orientation unknowns, and the sub-matrix $\boldsymbol{J}_i$ is the Jacobian matrix of the map projection determined for the $i$th new network point and created as follows:

$$\boldsymbol{J}_i = \begin{bmatrix} \dfrac{\partial \mathrm{e}}{\partial \lambda_i}(\lambda_i, \varphi_i) & \dfrac{\partial \mathrm{e}}{\partial \varphi_i}(\lambda_i, \varphi_i) \\ \dfrac{\partial \mathrm{n}}{\partial \lambda_i}(\lambda_i, \varphi_i) & \dfrac{\partial \mathrm{n}}{\partial \varphi_i}(\lambda_i, \varphi_i) \end{bmatrix} \tag{84}$$

The standard confidence ellipses of the new network points in the projected coordinate system can now rigorously be determined by using the covariance matrix elements from Eq. (82) and applying Eqs. (50) to (53).

In general, map projections used for horizontal geodetic network computation are given with rather complicated equations, and the corresponding Jacobian matrices are rarely published. However, for conformal projections, the scale factor and meridian convergence are always given. They can be derived by using only half of the elements of the Jacobian matrix, e.g. from partial derivatives of mapping equations with respect to the geodetic longitude (Vaníček and Krakiwsky 1986, pp. 360–361):

$$k_i = \frac{1}{N_i \cos \varphi_i} \sqrt{\left(\frac{\partial \mathrm{e}}{\partial \lambda_i}(\lambda_i, \varphi_i)\right)^2 + \left(\frac{\partial \mathrm{n}}{\partial \lambda_i}(\lambda_i, \varphi_i)\right)^2} \tag{85}$$

$$\gamma_i = \arctan\left(\frac{\partial \mathrm{n}}{\partial \lambda_i}(\lambda_i, \varphi_i) \Big/ \frac{\partial \mathrm{e}}{\partial \lambda_i}(\lambda_i, \varphi_i)\right) \tag{86}$$

The standard confidence ellipses in the projected coordinate system ($a_i$, $b_i$, and $t_i$) can also be obtained from the corresponding standard confidence ellipses in the local geodetic system ($\bar{a}_i$, $\bar{b}_i$, and $\bar{t}_i$), see Eqs. (50) to (53), by simply scaling and rotating them as follows:

$$a_i = k_i \bar{a}_i \quad b_i = k_i \bar{b}_i \quad t_i = \bar{t}_i - \gamma_i \tag{87)–(89}$$

where $k_i$ is the scale factor, and $\gamma_i$ is the meridian convergence at the $i$th new network point, see Eqs. (85) and (86). This is an alternative way to rigorously determine the standard confidence ellipse only in the case of a conformal mapping; otherwise, the corresponding elements of the Tissot's indicatrix – see e.g. Snyder (1987, pp. 20–27) – are required to be able to define the parameters of the affine transformation of the original standard confidence ellipse.

### *Nonrigorousness of Planar Adjustment Models*

Numerical examples of using the presented models for computation of classical horizontal geodetic networks in the geodetic and projected coordinate systems (in the following section) show that planar adjustment models provide equal and true results if but only if the observation set is error-free, see Table 2. The nonrigorousness of the planar adjustment model can be attributed to the stochastic modelling. It can be stressed that the first model (for the geodetic coordinate system) uses original observations which are – contrary to the conventional ellipsoidal model (cf. Krakiwsky and Thomson 1978; Vaníček and Krakiwsky 1986, pp. 401–403) – not reduced to the surface of the reference ellipsoid (Vincenty 1980).

The second model (for projected coordinate systems) – like the conventional planar adjustment model – uses observations reduced to the mapping plane. It can be expected that such adaptation of the original observations also changes their stochastic properties. An example is the introduction of quadrance and spread instead of the original distance and direction observations (Fuhrmann and Navratil 2013). The observations reduced to the mapping plane slightly change their stochastic properties and a substitute of the original weight matrix $\boldsymbol{P}$ for the reduced observations would be a fully populated inverse covariance matrix $\breve{\boldsymbol{P}}$. It should satisfy equations $\hat{\boldsymbol{x}}_{\boldsymbol{p}} = \mathrm{conv}_{\mathrm{e,n}}\, \hat{\boldsymbol{x}}_{\boldsymbol{g}}$ and $\boldsymbol{\Sigma}_{\hat{x}_p} = \boldsymbol{A_m} \boldsymbol{\Sigma}_{\hat{x}_g} \boldsymbol{A}_{\boldsymbol{m}}^{\mathrm{T}}$ which leads to:

$$\left[ \begin{array}{l} \left(\boldsymbol{A}_{\boldsymbol{p}}^{\mathrm{T}} \breve{\boldsymbol{P}} \boldsymbol{A}_{\boldsymbol{p}}\right)^{-1} \boldsymbol{A}_{\boldsymbol{p}}^{\mathrm{T}} \breve{\boldsymbol{P}} \boldsymbol{l} = \boldsymbol{0} \\ \boldsymbol{l}^{\mathrm{T}} \breve{\boldsymbol{P}} \boldsymbol{l} \left(\boldsymbol{A}_{\boldsymbol{p}}^{\mathrm{T}} \breve{\boldsymbol{P}} \boldsymbol{A}_{\boldsymbol{p}}\right)^{-1} = \boldsymbol{l}^{\mathrm{T}} \boldsymbol{P} \boldsymbol{l} \boldsymbol{A}_{\boldsymbol{m}} \left(\boldsymbol{A}_{\boldsymbol{g}}^{\mathrm{T}} \boldsymbol{P} \boldsymbol{A}_{\boldsymbol{g}}\right)^{-1} \boldsymbol{A}_{\boldsymbol{m}}^{\mathrm{T}} \end{array} \right. \tag{90}$$

Matrices $\boldsymbol{A_g}$, $\boldsymbol{A_p}$, $\boldsymbol{A_m}$, and vector $\boldsymbol{l}$ should be determined from the final coordinates of new network points estimated in the geodetic coordinate system (i.e., from the last iteration step) or from the corresponding coordinates converted to the projected coordinate system. However, the sought matrix $\breve{\boldsymbol{P}}$ is difficult to determine in this way. Matrices in the system of matrix Eqs. (90) are of dimensions $u \times 1$ and $u \times u$, respectively, with $u$ representing the number of network unknowns. Considering that weight and covariance matrices are symmetrical, this system provides $u(u+3)/2$ equations for $o(o+1)/2$ unknown inverse covariance matrix elements with $o$ representing the number of observations. Redundant observations (i.e., $o > u$) always lead to an underdetermined system of nonlinear (!) equations and a possible solution for $\breve{\boldsymbol{P}}$ is not unique.

Looking for a convenient way of determining an adequate inverse covariance matrix $\breve{\boldsymbol{P}}$ from Eq. (90) may be stored for possible future research. Obviously, much easier way to achieve a completely rigorous solution in the projected coordinate system is using mathematical model for the computation in the geodetic coordinate system. The results can further be transformed into the corresponding projected coordinate system in a rigorous way by using the mapping Eqs. (54) and (55), and Eq. (82).

## Numerical Examples

A fictitious network is created, and different map projections are used to test the performance of both mathematical models for computation of classical horizontal geodetic networks. This section presents:

- map projections used for testing,
- a fictitious test network,
- simulation of measurement campaigns, and
- results of the test computations with some remarks.

A self-developed computer program is used for testing. It is written in C++ and uses double-precision floating-point arithmetic. The matrix arithmetic algorithms are taken from Press et al. (1992, pp. 32–104).

### *Map Projections Used for Testing*

Three map projections from the reference ellipsoid to the plane are selected to test the proposed models for computation of classical horizontal geodetic networks in projected coordinate systems:

- Conformal Cylindrical (CC) projection,
- Equal-Area Cylindrical (EAC) projection, and
- Transverse Mercator (TM) projection.

The first two projections (CC and EAC) realize simple, rigorous mappings from a reference ellipsoid to the plane and are adapted for local applications (Safari and Ardalan 2007). The mapping equations for the Conformal Cylindrical projection are:

$$e_{CC_i} = N_0(\lambda_i - \lambda_0)\cos\varphi_0 \tag{91}$$

$$n_{CC_i} = N_0(\psi(\varphi_i) - \psi(\varphi_0))\cos\varphi_0 \tag{92}$$

where $\lambda_0$ and $\varphi_0$ are the selected standard longitude and latitude (i.e., a centroid of the network) and

$$\psi(\varphi) = \operatorname{arcsinh}(\tan\varphi) - e\operatorname{arctanh}(e\sin\varphi) \tag{93}$$

is the isometric latitude (Snyder 1987, p. 15). The corresponding Jacobian matrix for the CC projection – to be able to create the mapping design matrix, Eq. (83), – is (Safari and Ardalan 2007):

$$\boldsymbol{J}_{CC_i} = \begin{bmatrix} N_0\cos\varphi_0 & 0 \\ 0 & \dfrac{M_i N_0\cos\varphi_0}{N_i\cos\varphi_i} \end{bmatrix} \tag{94}$$

The scale factor and meridian convergence for the CC projection are:

$$k_{CC_i} = \frac{N_0\cos\varphi_0}{N_i\cos\varphi_i} \quad \gamma_{CC_i} = 0 \tag{95)–(96}$$

The mapping equations for the Equal-Area Cylindrical projection are (Safari and Ardalan 2007):

$$e_{EAC_i} = N_0(\lambda_i - \lambda_0)\cos\varphi_0 \tag{97}$$

$$n_{EAC_i} = \frac{a^2}{2N_0\cos\varphi_0}(\mathrm{q}(\varphi_i) - \mathrm{q}(\varphi_0)) \tag{98}$$

where $\lambda_0$ and $\varphi_0$ are the selected standard longitude and latitude and

$$\mathrm{q}(\varphi) = (1 - e^2)\left(\frac{\sin\varphi}{1 - e^2\sin^2\varphi} + \frac{\operatorname{arctanh}(e\sin\varphi)}{e}\right) \tag{99}$$

is an auxiliary parameter used to define the authalic latitude (Snyder 1987, p. 101). The corresponding Jacobian matrix for the EAC projection is (Safari and Ardalan 2007):

$$\boldsymbol{J}_{EAC_i} = \begin{bmatrix} N_0 \cos\varphi_0 & 0 \\ 0 & \dfrac{M_i N_i \cos\varphi_i}{N_0 \cos\varphi_0} \end{bmatrix} \tag{100}$$

Since the EAC projection is not conformal, the scale factor changes with the azimuth ($A$). However, the extreme scale factors at the $i$th network point – the major and minor semi-axis of its Tissot's indicatrix – can be derived (e.g. Snyder 1987, pp. 20–27). The obtained maximum and minimum scale factors for the EAC projection are as follows:

$$\max_{0 \le A < 2\pi} k_{EAC_i} = \frac{k_i^2 + 1 + |k_i^2 - 1|}{2k_i} \tag{101}$$

$$\min_{0 \le A < 2\pi} k_{EAC_i} = \frac{k_i^2 + 1 - |k_i^2 - 1|}{2k_i} \tag{102}$$

where $k_i$ is equal to the scale factor for the CC projection ($k_{CC_i}$), see Eq. (95).

The inverse mapping for the CC and EAC projections is realized with the Newton-Raphson iteration method as follows (Bildirici 2017):

$$d_i = \frac{\partial \mathrm{e}}{\partial \lambda_i}(\lambda_i, \varphi_i) \frac{\partial \mathrm{n}}{\partial \varphi_i}(\lambda_i, \varphi_i) - \frac{\partial \mathrm{e}}{\partial \varphi_i}(\lambda_i, \varphi_i) \frac{\partial \mathrm{n}}{\partial \lambda_i}(\lambda_i, \varphi_i) \tag{103}$$

$$\tilde{\lambda}_i = \lambda_i + \frac{1}{d_i}\left( (\mathrm{n}(\lambda_i, \varphi_i) - n_i) \frac{\partial \mathrm{e}}{\partial \varphi_i}(\lambda_i, \varphi_i) - (\mathrm{e}(\lambda_i, \varphi_i) - e_i) \frac{\partial \mathrm{n}}{\partial \varphi_i}(\lambda_i, \varphi_i) \right) \tag{104}$$

$$\tilde{\varphi}_i = \varphi_i + \frac{1}{d_i}\left( (\mathrm{e}(\lambda_i, \varphi_i) - e_i) \frac{\partial \mathrm{n}}{\partial \lambda_i}(\lambda_i, \varphi_i) - (\mathrm{n}(\lambda_i, \varphi_i) - n_i) \frac{\partial \mathrm{e}}{\partial \lambda_i}(\lambda_i, \varphi_i) \right) \tag{105}$$

where $d_i$ is the Jacobian determinant, i.e., the determinant of the Jacobian matrix in Eq. (84), and $(\tilde{\lambda}_i, \tilde{\varphi}_i)$ are the improved geodetic coordinates (to be used in the next iteration step) of the $i$th network point, which is given with the projected coordinates $(e_i, n_i)$.

Exact mapping equations for the Transverse Mercator projection could be implemented by using Jacobian elliptic functions (Lee 1976). However, the extensions of Krüger's series are used here. The accuracy of a few nanometers within the 3,900 km of the central meridian is guaranteed, which is comparable with the exact method (Karney 2011). The mapping equations for $e_{TM_i}$ and $n_{TM_i}$ based on the Karney series as well as the highly accurate scale factor $k_{TM_i}$ and the meridian convergence $\gamma_{TM_i}$ published by Kawase (2013) are not repeated here.

The Conformal Cylindrical and Equal-Area Cylindrical projections are selected because of their simple, closed-form direct mapping equations and the corresponding Jacobian matrices. This is convenient for testing purposes – to obtain a solution of a horizontal geodetic network in the projected coordinate system without loss of accuracy. The Transverse Mercator projection is selected as an example of mapping used for the computation of classical horizontal geodetic networks worldwide.

The reversibility check for all three projections (TM, CC, and EAC) is performed for the test network point coordinates (see below). The maximum positional inaccuracy after conversions from the projected to the geodetic and back to the projected coordinates amounts to 0.000000002 m (2 nm) for the TM projection. For the CC and EAC projections no differences are detected (i.e., the machine precision is achieved).

***Fictitious Test Network***

The test network consists of six existing mountain peaks located in six European countries, see Fig. 2.

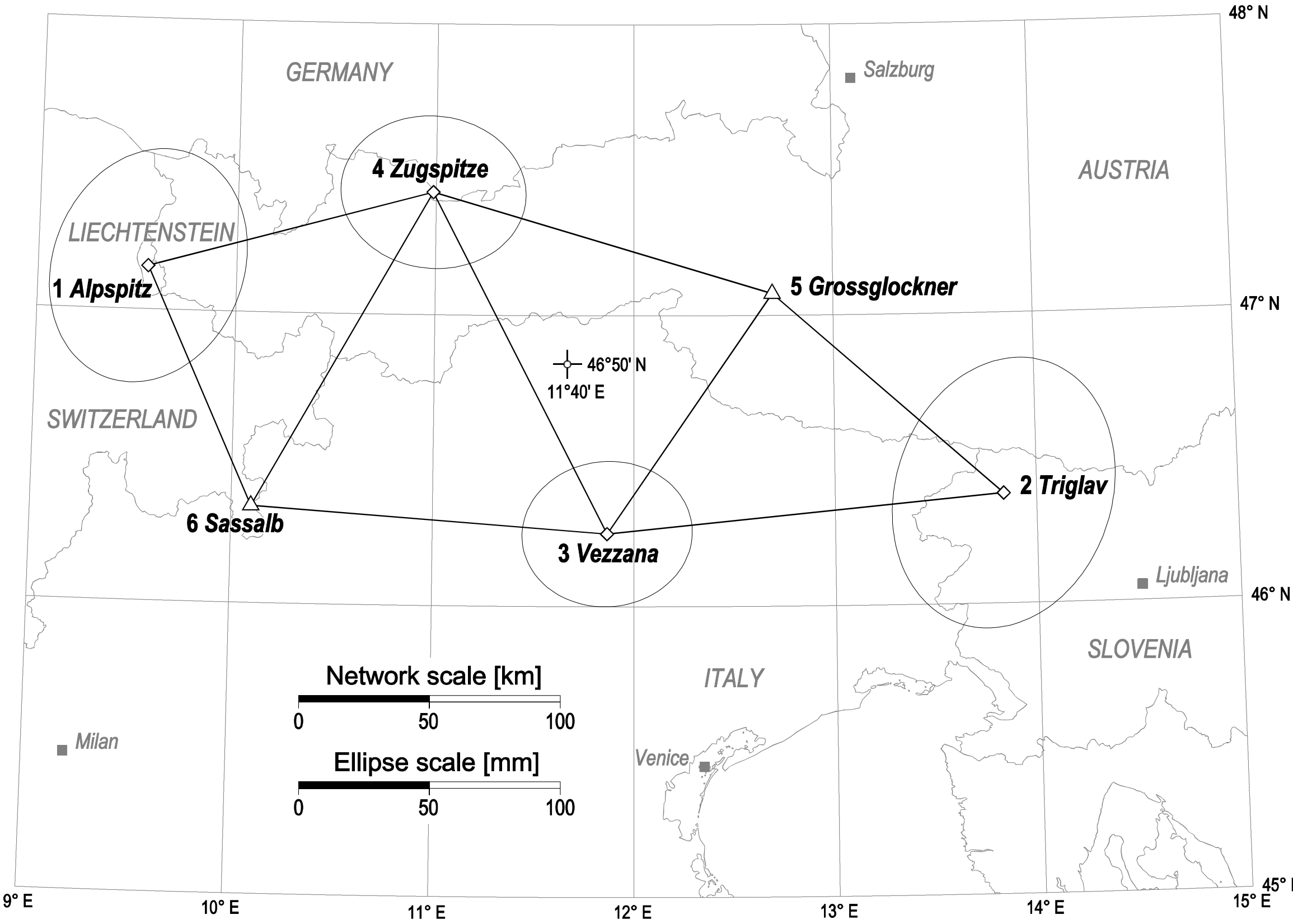

***Fig. 2.*** *The network for testing the performance of the proposed mathematical models*

The distances between adjacent network points range from 100 km to 152 km, which is far beyond the maximum range of classical geodetic measurements. The selected geodetic coordinates of the test network points are taken from Wikipedia, see Table 1; the longitudes and latitudes of the selected mountain peaks are rounded to arcseconds, and the original heights (above sea level) are assumed to be ellipsoidal for simplicity.

**Table 1.** Geodetic coordinates (exact) and locations of the test network points

| Pt. № | $\lambda$ [dms] | $\varphi$ [dms] | $h$ [m] | Mountain Peak Name |
|---|---|---|---|---|
| 1 | 9°33′14″E | 47°08′55″N | 1934 | Alpspitz, Liechtenstein |
| 2 | 13°50′12″E | 46°22′42″N | 2864 | Triglav, Slovenia |
| 3 | 11°52′02″E | 46°15′00″N | 3192 | Vezzana, Italy |
| 4 | 10°59′07″E | 47°25′16″N | 2962 | Zugspitze, Germany |
| 5 | 12°41′43″E | 47°04′30″N | 3798 | Grossglockner, Austria |
| 6 | 10°05′56″E | 46°20′02″N | 2862 | Sassalb, Switzerland |

Two of the network points (5 and 6) are network points with known (fixed) coordinates that define the geodetic datum of the network. The other network points (1 to 4) are points with unknown (newly determined) coordinates. The heights of all network points are considered fixed (constant values). There are (see also Fig. 2):

- 14 network unknowns (eight horizontal coordinates and six orientation angles) and
- 27 observations (nine distances and twice as many directions).

The GRS80 reference ellipsoid is used with the parameters (Moritz 2000):

- $a$ = 6378137 m and
- $e^2$ = 0.0066943800229.

The parameters of the projections (TM, CC, and EAC) used for the tests are adapted to the location of the network. The parameters of the TM projection are:

- $\lambda_0$ = 12°E (central meridian),
- $k_0$ = 0.9998 (scale factor at the central meridian),
- $e_0$ = 500000 m (false easting), and
- $n_0$ = −5000000 m (false northing).

The estimated network centroid is used to define the parameters of the CC and EAC projections (see Fig. 2):

- $\lambda_0$ = 11°40′E (standard longitude) and
- $\varphi_0$ = 46°50′N (standard latitude).

It should be noted that the CC and EAC projections lead to much larger distortions in the scale than the TM projection. Applying the above-mentioned parameters of the map projections, the scale factors obtained in the fictitious test network points vary between 0.99980 and 1.00022 (distortions between −0.22‰ and +0.20‰) for the TM projection, between 0.98935 and 1.01108 (distortions between −11.08‰ and +10.65‰) for the CC projection, see Eq. (95), and between 0.98904 and 1.01108 (distortions between −11.08‰ and +10.96‰) for the EAC projection, see Eqs. (101) and (102). Also, an experiment with geodetic area calculations based on the well-known Lambert Equal-Area Cylindrical projection of the world (e.g. Snyder 1987, p. 81–85), which is undistorted along the equator, and a regionally adapted EAC projection used in this work, was carried out at similar latitudes as the test network in Fig. 2. Somehow surprisingly, the obtained areas for both (i.e., world and regionally adapted) equal-area map projections suffer from large inaccuracies of the same order of magnitude (Berk and Ferlan 2018). These inaccuracies are caused by large differences between the projected geodesics and the chords connecting the polygon vertices, which also implies large arc-to-chord reductions of geodetic observations.

### *Simulation of Measurement Campaigns*

The measurement campaigns are simulated in two different ways. The observations are generated as:

- error-free observations and
- error-prone observations.

The error-free observations are calculated from the exact network coordinates given in Table 1 using Eqs. (14) and (15). The error-prone observations are generated by:

- rounding the error-free distances to the nearest even decimeter (errors up to ±10 cm) and
- rounding the error-free directions in decimal degrees to four digits (errors up to ±0.18″).

The corresponding a priori standard deviations are set to:

- ±6.9 cm for all distance observations and
- ±0.11″ for all direction observations.

These are the rounded RMS values of the error-prone observations generated as described above. The observations are treated as uncorrelated, and the weights of observations are determined as reciprocals of their a priori variances.

The specified input data are sufficient to ensure the repeatability of the numerical tests without ambiguity. In the test computations, the initial approximate values of the network unknowns are generated by:

- rounding the exact geodetic coordinates, see Table 1, in decimal degrees to two digits (errors up to ±18″, corresponding to about ±500 m) and
- rounding the calculated (error-free) orientation angles in decimal degrees to two digits (errors up to ±18″).

However, these initial values only affect intermediate results (e.g. Table 2) and not the final results of computations.

### *Computations of the Test Network with Error-Free Observations*

The first computational experiment is performed using the error-free observation set. Both mathematical models are tested: in the geodetic and projected coordinate systems – the latter by applying the TM, CC, and EAC projections. Table 2 shows inaccuracies expressed in terms of the maximum positional errors at the new network points (i.e., displacements from their exact positions): $\epsilon_p = \max_i \epsilon_{p_i}$. The errors refer to the computation in the geodetic coordinate system ($\epsilon_{p_G}$) and in the TM-, CC-, and EAC-projection-based coordinate systems ($\epsilon_{\tilde{p}_{TM}}$, $\epsilon_{\tilde{p}_{CC}}$, and $\epsilon_{\tilde{p}_{EAC}}$).

**Table 2.** Maximum positional errors at the new network points in the iterative network computation by using the error-free observation set

| Iteration | $\epsilon_{p_G}$ [m] | $\epsilon_{\tilde{p}_{TM}}$ [m] | $\epsilon_{\tilde{p}_{CC}}$ [m] | $\epsilon_{\tilde{p}_{EAC}}$ [m] |
|---|---|---|---|---|
| 1 | 0.900503662 | 0.867203492 | 3.410963397 | 5.354982986 |
| 2 | 0.000002672 | 0.000216349 | 0.037537851 | 0.063869553 |
| 3 | 0.000000001 | 0.000000065 | 0.000602163 | 0.000965107 |
| 4 | 0.000000000 | 0.000000002 | 0.000006511 | 0.000013868 |
| 5 | 0.000000000 | 0.000000003 | 0.000000061 | 0.000000216 |
| 6 | 0.000000000 | 0.000000002 | 0.000000002 | 0.000000003 |
| 7 | 0.000000000 | 0.000000001 | 0.000000001 | 0.000000001 |
| 8 | 0.000000000 | 0.000000003 | 0.000000000 | 0.000000001 |

The maximum positional errors in Table 2 result from the exact and calculated geodetic coordinates using Eq. (14), which eliminates the influence of the scale factor. The necessary conversion from the projected coordinates (TM, CC, and EAC) is performed using the corresponding inverse mapping equations. In all computations, a positional accuracy in the nanometer range is achieved, except for the TM-projection-based coordinate system, where the maximum positional errors oscillate between 1 nm and 3 nm, which is obviously due to the afore-mentioned limited reversibility (to about 2 nm) of the Karney series for the TM projection.

The experiment with the error-free observations confirms that both functional models (for the geodetic and projected coordinate systems) provide equal and true results – with an accuracy of a few nanometers.

### *Computations of the Test Network with Error-Prone Observations*

The second experiment is performed using the simulated error-prone observation set. Comparisons are given between the results of computation in the geodetic coordinate system, which are rigorously converted into the projected coordinate systems, and the results of planar network computations based on the TM, CC, and EAC projections. Tables 3, 5, and 7 show rigorously estimated pairs of coordinates of new network points ($e_i, n_i$) from the computation in the geodetic coordinate system, which are further converted by using the corresponding mapping equations. The coordinate errors $\epsilon_{\tilde{e}_i}$ and $\epsilon_{\tilde{n}_i}$ in the planar network computation are also given, which are defined as follows:

$$\epsilon_{\tilde{e}_i} = e_i - \tilde{e}_i \quad \epsilon_{\tilde{n}_i} = n_i - \tilde{n}_i \qquad (106)\text{–}(107)$$

with ($\tilde{e}_i, \tilde{n}_i$) as the corresponding pairs of coordinates determined in a planar network computation.

In addition, comparisons of the network coordinate accuracy estimates (standard confidence ellipses) are given. Tables 4, 6, and 8 show the rigorously estimated elements of standard confidence ellipses of new network points ($a_i$, $b_i$, and $t_i$) from the computation in the geodetic coordinate system. For the CC- and EAC-projection-based coordinate systems, Eqs. (51) to (53) are applied using the corresponding covariance matrix from Eq. (82). For the CC- and TM-projection-based coordinate systems, Eqs. (87) to (89) are applied using the corresponding covariance matrix in linear units from Eq. (47). For the CC-projection-based coordinate system, equal results are obtained by using both approaches. The errors of the standard confidence ellipse elements $\epsilon_{\tilde{a}_i}$, $\epsilon_{\tilde{b}_i}$, and $\epsilon_{\tilde{t}_i}$ in the planar network computation are also given, which are defined as follows:

$$\epsilon_{\tilde{a}_i} = a_i - \tilde{a}_i \quad \epsilon_{\tilde{b}_i} = b_i - \tilde{b}_i \quad \epsilon_{\tilde{t}_i} = t_i - \tilde{t}_i \qquad (108)\text{–}(110)$$

with $\tilde{a}_i$, $\tilde{b}_i$, and $\tilde{t}_i$ as the corresponding elements of the standard confidence ellipses in a planar network computation.

**Table 3.** Coordinates of points in the TM-projection-based coordinate system ($e_{TM}$, $n_{TM}$) and their errors in the planar computational model ($\epsilon_{\tilde{e}_{TM}}$, $\epsilon_{\tilde{n}_{TM}}$)

| Pt. № | $e_{TM}$ | $n_{TM}$ | $\epsilon_{\tilde{e}_{TM}}$ [m] | $\epsilon_{\tilde{n}_{TM}}$ [m] |
|---|---|---|---|---|
| 1 | 314516.322644 | 225627.201222 | 0.000000 | 0.000008 |
| 2 | 641272.110250 | 138751.296733 | 0.000012 | 0.000003 |
| 3 | 489763.038340 | 122858.144890 | 0.000012 | 0.000000 |
| 4 | 423448.373783 | 253512.338335 | 0.000000 | 0.000008 |
| Ext | — | — | 0.000012 | 0.000008 |

The coordinate errors in the planar adjustment of the test network based on the Transverse Mercator projection, see Table 3, reach up to 0.012 mm.

**Table 4.** Elements of the standard confidence ellipses of points in the TM-projection-based coordinate system ($a_{TM}$, $b_{TM}$, $t_{TM}$) and their errors in the planar computational model ($\epsilon_{\tilde{a}_{TM}}$, $\epsilon_{\tilde{b}_{TM}}$, $\epsilon_{\tilde{t}_{TM}}$)

| Pt. № | $a_{TM}$ [m] | $b_{TM}$ [m] | $t_{TM}$ [dms] | $\epsilon_{\tilde{a}_{TM}}$ [m] | $\epsilon_{\tilde{b}_{TM}}$ [m] | $\epsilon_{\tilde{t}_{TM}}$ [dms] |
|---|---|---|---|---|---|---|
| 1 | 0.045717 | 0.036396 | 21°46′09″ | −0.000001 | 0.000001 | 7″ |
| 2 | 0.052758 | 0.041291 | 18°26′35″ | −0.000004 | −0.000010 | 1′04″ |
| 3 | 0.032552 | 0.027737 | 85°01′47″ | −0.000012 | −0.000005 | 1′38″ |
| 4 | 0.035402 | 0.029095 | 95°46′13″ | −0.000013 | −0.000005 | 57″ |
| Ext | 0.052758 | — | — | −0.000013 | −0.000010 | 1′38″ |

The inaccuracies in the standard confidence ellipse elements in the adjustment of the test network based on the Transverse Mercator projection, see Table 4, reach up to 0.013 mm for the semi-axes and up to 1′38″ for their azimuths. Relative errors in the semi-axes of the standard confidence ellipses (e.g. $\epsilon_{\tilde{a}_i}/\tilde{a}_i$) up to 0.36‰ are obtained.

**Table 5.** Coordinates of points in the CC-projection-based coordinate system ($e_{CC}$, $n_{CC}$) and their errors in the planar computational model ($\epsilon_{\tilde{e}_{CC}}$, $\epsilon_{\tilde{n}_{CC}}$)

| Pt. № | $e_{CC}$ | $n_{CC}$ | $\epsilon_{\tilde{e}_{CC}}$ [m] | $\epsilon_{\tilde{n}_{CC}}$ [m] |
|---|---|---|---|---|
| 1 | −161188.419322 | 35152.648583 | −0.000089 | 0.000366 |
| 2 | 165554.075154 | −50367.595878 | −0.000040 | −0.000286 |
| 3 | 15300.795003 | −64497.267106 | −0.000197 | −0.000042 |
| 4 | −51984.672290 | 65705.176800 | −0.000066 | 0.000106 |
| Ext | — | — | −0.000197 | 0.000366 |

The coordinate errors in the planar adjustment of the test network based on the Conformal Cylindrical projection, see Table 5, reach up to 0.366 mm.

**Table 6.** Elements of the standard confidence ellipses of points in the CC-projection-based coordinate system ($a_{CC}$, $b_{CC}$, $t_{CC}$) and their errors in the planar computational model ($\epsilon_{\tilde{a}_{CC}}$, $\epsilon_{\tilde{b}_{CC}}$, $\epsilon_{\tilde{t}_{CC}}$)

| Pt. № | $a_{CC}$ [m] | $b_{CC}$ [m] | $t_{CC}$ [dms] | $\epsilon_{\tilde{a}_{CC}}$ [m] | $\epsilon_{\tilde{b}_{CC}}$ [m] | $\epsilon_{\tilde{t}_{CC}}$ [dms] |
|---|---|---|---|---|---|---|
| 1 | 0.045977 | 0.036603 | 19°58′31″ | 0.000208 | 0.000166 | 30″ |
| 2 | 0.052315 | 0.040944 | 19°46′22″ | −0.000273 | −0.000296 | 36′06″ |
| 3 | 0.032211 | 0.027447 | 84°56′01″ | −0.000254 | −0.000154 | 11′40″ |
| 4 | 0.035799 | 0.029421 | 95°01′23″ | 0.000279 | 0.000247 | 5′14″ |
| Ext | 0.052315 | — | — | 0.000279 | −0.000296 | 36′06″ |

The inaccuracies in the standard confidence ellipse elements in the adjustment of the test network based on the Conformal Cylindrical projection, see Table 6, reach up to 0.296 mm for the semi-axes and up to 36′06″ for their azimuths. Relative errors in the semi-axes of the standard confidence ellipses up to 8.39‰ are obtained.

**Table 7.** Coordinates of points in the EAC-projection-based coordinate system ($e_{EAC}$, $n_{EAC}$) and their errors in the planar computational model ($\epsilon_{\tilde{e}_{EAC}}$, $\epsilon_{\tilde{n}_{EAC}}$)

| Pt. № | $e_{EAC}$ | $n_{EAC}$ | $\epsilon_{\tilde{e}_{EAC}}$ [m] | $\epsilon_{\tilde{n}_{EAC}}$ [m] |
|---|---|---|---|---|
| 1 | −161188.419322 | 34946.914738 | −0.000226 | 0.000327 |
| 2 | 165554.075154 | −50792.210747 | −0.000013 | −0.000331 |
| 3 | 15300.795003 | −65194.134741 | −0.000186 | −0.000034 |
| 4 | −51984.672290 | 64987.791999 | −0.000146 | 0.000082 |
| Ext | — | — | −0.000226 | −0.000331 |

The coordinate errors in the planar adjustment of the test network based on the Equal-Area Cylindrical projection, see Table 7, reach up to 0.331 mm.

**Table 8.** Elements of the standard confidence ellipses of points in the EAC-projection-based coordinate system ($a_{EAC}$, $b_{EAC}$, $t_{EAC}$) and their errors in the planar computational model ($\epsilon_{\tilde{a}_{EAC}}$, $\epsilon_{\tilde{b}_{EAC}}$, $\epsilon_{\tilde{t}_{EAC}}$)

| Pt. № | $a_{EAC}$ [m] | $b_{EAC}$ [m] | $t_{EAC}$ [dms] | $\epsilon_{\tilde{a}_{EAC}}$ [m] | $\epsilon_{\tilde{b}_{EAC}}$ [m] | $\epsilon_{\tilde{t}_{EAC}}$ [dms] |
|---|---|---|---|---|---|---|
| 1 | 0.045505 | 0.036550 | 20°58′47″ | −0.000301 | 0.000060 | 34′06″ |
| 2 | 0.053103 | 0.041018 | 18°33′56″ | 0.000462 | −0.000229 | −40′02″ |
| 3 | 0.032217 | 0.028036 | 84°09′03″ | −0.000296 | 0.000396 | −50′03″ |
| 4 | 0.035793 | 0.028784 | 94°30′39″ | 0.000307 | −0.000366 | −25′56″ |
| Ext | 0.053103 | — | — | 0.000462 | 0.000396 | −50′03″ |

The inaccuracies in the standard confidence ellipse elements in the adjustment of the test network based on the Equal-Area Cylindrical projection, see Table 8, reach up to 0.462 mm for the semi-axes and up to 50′03″ for their azimuths. Relative errors in the semi-axes of the standard confidence ellipses up to 14.13‰ are obtained.

The experiment with the error-prone observations confirms the assumption regarding the nonrigorous stochastic model of the conventional planar network adjustment. It also clearly shows that minimizing the distortions in the scale of mapping increases the computational accuracy. However, the use of an optimal map projection for the network area is a well-known recommendation (e.g. Kuang 1996, p. 59) that avoids large differences between the real-world and map-grid dimensions of geographic phenomena. An approach addressing this problem by minimizing ground-to-grid distortions was recently presented by Baselga (2021). This is particularly important in civil engineering applications – to integrate CAD and GIS data environments (e.g. Habib et al. 2019).

## Further Remarks on the Rigorousness of the Proposed Planar Adjustment Model

Aiming to present the ideas in the paper as clear as possible, the assumption of mark-to-mark corrected observations is used as a starting point. However, the preliminary mark-to-mark corrections of the distance and direction observations from the $i$th standpoint to the $j$th forepoint can easily be avoided. One can simply replace $h_i$ with $h_i + \mathscr{i}_i$ in Eq. (13) and $h_j$ with $h_j + \mathscr{t}_j$ in Eqs. (11) to (13), (28), (30) to (32), and (34) to (36), where $\mathscr{i}_i$ is the height of the surveying instrument at the standpoint (e.g. the height of the optical center of the total station above the top of the survey mark) and $\mathscr{t}_j$ is the height of the target at the forepoint (e.g. the height of the optical center of the retroreflector prism above the top of the survey mark). The mark-to-mark correction of the direction observations is very small and the height of the instrument at the standpoint is not involved in this correction. Different targets (placed at different heights) can be used for a pair of distance and direction observations ($D'_{i,j}$ and $R'_{i,j,k}$) for a particular network connection ($i \longrightarrow j$).

Also, the preliminary transformation of direction observations from the local astronomic to the local geodetic system can be avoided. One can start from the Laplace's equation of orientation (e.g. Vaníček and Krakiwsky 1986, p. 348):

$$\bar{A}_{i,j} - A_{i,j} = \eta_i \tan \varphi_i + \left(\xi_i \sin A_{i,j} - \eta_i \cos A_{i,j}\right) \cot Z_{i,j} \tag{111}$$

and express the astronomic azimuth in terms of the network unknowns by applying Eqs. (5) to (7) as:

$$\bar{A}_{i,j} = A_{i,j} + \eta_i \tan \varphi_i + \varDelta z_{i,j} \left(\frac{\xi_i \sin^2 A_{i,j}}{\varDelta y_{i,j}} - \frac{\eta_i \cos^2 A_{i,j}}{\varDelta x_{i,j}}\right) \tag{112}$$

where $\bar{A}_{i,j}$ is the astronomic azimuth from the $i$th standpoint to the $j$th forepoint, $A_{i,j}$ is the corresponding geodetic azimuth from Eq. (15), $\varDelta x_{i,j}$, $\varDelta y_{i,j}$, and $\varDelta z_{i,j}$ are coordinate differences from Eqs. (11) to (13), $\xi_i$ is the meridian component and $\eta_i$ is the prime vertical component of the vertical deflection. These components should not be confused with the auxiliary parameters $\xi_{i,j}$ and $\eta_{i,j}$ which are defined by Eqs. (24) and (27). The sign of both deflection components is positive if the vertical is farther north and farther east than the normal for both the northern and southern hemispheres (e.g. Vaníček and Krakiwsky 1986, p. 93).

Rigorous equations for determining the differential variations of the vertical deflections and the geodetic azimuth as a function of the changes in geodetic coordinates were presented by Soler et al. (2014). But the heights of the instruments and targets and the components of the vertical deflections in the classical horizontal geodetic networks are normally considered as auxiliary observations (i.e., as constant values) and the proposed planar functional model can easily be adapted. One can use the mark-to-mark corrected direction observations in the local astronomic system instead of those in the local geodetic system and replace $A_{i,j}$ – i.e., $\operatorname{arctan2g}\left(\varDelta y_{i,j}, \varDelta x_{i,j}\right)$ – in Eq. (70) with the right side of Eq. (112). The mark-to-mark corrected distance observations are not affected by the vertical deflections.

The fact is that the proposed solution, which considers the heights of the instruments/targets and the vertical deflections, assumes that the instruments and targets share their horizontal geodetic coordinates $(\lambda, \varphi)$ with the corresponding survey marks (in contradiction to the phenomena considered). Their heights should be measured along the normal to the ellipsoid. The coordinate differences between an instrument at the $i$th standpoint and the corresponding survey mark can be determined in three steps as follows:

$$\varDelta z_i = \imath_i / \sqrt{1 + \tan^2 \xi_i + \tan^2 \eta_i} \tag{113}$$

$$\varDelta y_i = \varDelta z_i \tan \eta_i \tag{114}$$

$$\varDelta x_i = \varDelta z_i \tan \xi_i \tag{115}$$

These coordinate differences should not be confused with the coordinate differences between the $i$th standpoint and the $j$th forepoint which are defined by Eqs. (11) to (13). According to Wikipedia, the largest vertical deflections in Central Europe can be found near the Grossglockner peak, see Fig. 2; the approximate values are +50″ for $\xi_i$ and −30″ for $\eta_i$. The inaccuracies caused by ignoring the vertical deflection for an instrument placed 2.0 m above the survey mark ($\imath_i$) would be as follows: 0.48 mm for the north ($\varDelta x_i$), −0.29 mm for the east ($\varDelta y_i$), and −0.08 μm for the up component ($\varDelta z_i - \imath_i$). Vertical deflections in rather flat areas are usually up to 15″, which would result in the coordinate errors up to 0.15 mm.

To create a completely rigorous functional model of horizontal geodetic network adjustment, the coordinate differences in Eqs. (113) to (115) should also be taken into account. One can obtain the 3D Cartesian coordinates of the survey mark $(X_i, Y_i, Z_i)$ from its geodetic coordinates (e.g. Ghilani and Wolf 2006, p. 317) as follows:

$$X_i = (N_i + h_i) \cos \varphi_i \cos \lambda_i \tag{116}$$

$$Y_i = (N_i + h_i) \cos \varphi_i \sin \lambda_i \tag{117}$$

$$Z_i = (N_i(1 - e^2) + h_i) \sin \varphi_i \tag{118}$$

The coordinate differences in Eqs. (113) to (115) can be converted from the local geodetic system to the 3D Cartesian coordinate system by applying the corresponding rotation matrix. The converted coordinate differences can be added to the coordinates of the survey mark from Eqs. (116) to (118) as follows:

$$\bar{X}_i = X_i - \varDelta y_i \sin\lambda_i - \cos\lambda_i(\varDelta x_i \sin\varphi_i - \varDelta z_i \cos\varphi_i) \tag{119}$$

$$\bar{Y}_i = Y_i + \varDelta y_i \cos\lambda_i - \sin\lambda_i(\varDelta x_i \sin\varphi_i - \varDelta z_i \cos\varphi_i) \tag{120}$$

$$\bar{Z}_i = Z_i + \varDelta x_i \cos\varphi_i + \varDelta z_i \sin\varphi_i \tag{121}$$

The obtained 3D Cartesian coordinates $(\bar{X}_i, \bar{Y}_i, \bar{Z}_i)$ refer to the instrument placed at the survey mark. They can be easily incorporated into the proposed planar network adjustment model. In each iteration step, the estimated planar coordinates of a network point $(e_i, n_i)$ should be converted to the geodetic coordinates $(\lambda_i, \varphi_i)$ by using the corresponding inverse mapping equations. Considering the known ellipsoidal height $(h_i)$, they should be further converted into the 3D Cartesian coordinates of the survey mark $(X_i, Y_i, Z_i)$ using Eqs. (116) to (118). Also considering the height of the instrument $(\imath_i)$ and the components of the vertical deflection $(\xi_i, \eta_i)$, the 3D Cartesian coordinates of the instrument $(\bar{X}_i, \bar{Y}_i, \bar{Z}_i)$ can be obtained using Eqs. (119) to (121). The latter should be converted back to the geodetic coordinates $(\bar{\lambda}_i, \bar{\varphi}_i, \bar{h}_i)$ using one of the various exact methods (e.g. Borkowski 1989; Zhang et al. 2005; Sjöberg 2008; Vermeille 2002, 2004, and 2011). The corresponding geodetic coordinates of the target $(\bar{\lambda}_j, \bar{\varphi}_j, \bar{h}_j)$ can be determined analogously. The obtained geodetic coordinates of the instruments/targets can be used to determine the misclosures in Eqs. (69) and (70), considering the replacement of the geodetic azimuth $(A_{i,j})$ with the astronomic azimuth $(\bar{A}_{i,j})$ from Eq. (112); see above. To determine the observables in Eqs. (8) to (13), the aforementioned geodetic coordinates of instruments $(\bar{\lambda}_i, \bar{\varphi}_i, \bar{h}_i)$ and targets $(\bar{\lambda}_j, \bar{\varphi}_j, \bar{h}_j)$ should be used instead of the geodetic coordinates of survey marks $(\lambda_i, \varphi_i, h_i)$ and $(\lambda_j, \varphi_j, h_j)$, respectively.

One should have in mind that the deflections of the vertical at the survey mark and the instrument above it are considered equivalent, which can be assumed for all practical purposes (e.g. Soler et al. 2014).

## Discussion and Conclusions

Two rigorous functional models for adjustment of classical horizontal geodetic networks are investigated – for computations in the geodetic and projected coordinate systems. Both are based on the parametric model of the three-dimensional geodetic network adjustment using geodetic coordinates $(\lambda, \varphi, h)$. The height-controlled approach is used; the ellipsoidal heights of network points are fixed – they should be determined beforehand. The mark-to-mark corrected observations are used as a starting point. The first model is completely rigorous and serves as a reference in the study. The second model is based on the conventional computational approach with the planar network adjustment using projected coordinates $(e, n)$. A strict distinction between the observation (preprocessed measured value) and the estimated observable (its most likely value) is maintained. In the proposed computational procedures, the observations are used exclusively for the determination of the misclosure vectors, see Eqs. (18), (19), (69), and (70). For computation in projected coordinate systems, each iteration step involves:

- conversion of the projected coordinates into the geodetic coordinates, Eqs. (56) and (57), and
- closed-form geometric reduction of mark-to-mark corrected observations directly to the mapping plane, Eqs. (63) and (64), instead of the classical preliminary stepwise reduction.

The geometric reductions mentioned above do not actually have to be carried out explicitly. One can simply determine the misclosure vector in the same way as for the computation in the geodetic coordinate system, see Eqs. (69) and (70).

When using an error-free observation set, both mathematical models yield equal and true coordinates of the network points (Table 2). However, if one follows the conventional computational approach of reducing terrestrial observations to the mapping plane the equality of both solutions is lost when using an error-prone observation set. This can be attributed to the nonrigorous stochastic model of the conventional planar network adjustment approach, and the need to adapt the weight matrix of the original observations is indicated, see Eq. (90). However, to obtain a completely rigorous adjustment model (functional and stochastic parts) in the projected coordinate system, the rigorous computation in the geodetic coordinate system is a more convenient approach. The projected coordinates can be easily determined from the geodetic coordinates using the corresponding mapping equations. The rigorous accuracy estimates can be determined according to the law of variance-covariance propagation using the mapping design matrix – Eq. (83).

The impact of nonrigorous consideration of the stochastic properties of the reduced observations on the resulting coordinates of the network points and their accuracy estimates are very small. Obviously, the obtained inaccuracies depend on the measurement accuracy (e.g. error-free results from error-free observation set) and on the distortions in the scale of mapping (compare distortions of the TM with the CC and EAC projections). The numerical example with very long network sides (see Fig. 2) detects coordinate errors and errors in the semi-axes of the standard confidence ellipses that are smaller than 0.5 mm for the CC and EAC projections, while for the TM-projection these errors are smaller than 0.02 mm. In the surveying practice, the limited ability of a priori accuracy estimation of observations (including covariances) may have a much larger impact on the estimated network unknowns. The presented planar model of horizontal geodetic network adjustment meets the requirements for processing the most accurate geodetic networks, regardless of the network size and the network point displacements. It can be particularly useful for applications in engineering surveys.

In classical geodetic literature, conformal mapping is an assumption in horizontal geodetic network computation. The proposed functional model of the planar network adjustment has no limitations regarding the properties of map projections and no extra effort is needed to correctly perform ground-to-grid reductions of observations. It could be adapted to a triaxial ellipsoid and map projections from the latter. Also, an appropriately adapted functional model can be rigorous in dealing with the deflections of the vertical, see Eqs. (112) to (121).

The presented rigorous functional model of the planar network adjustment is very simple; it is realized by using Eqs. (3), (8) to (15), (60), and (69) to (79). The only price to pay – as compared to the computation in the geodetic coordinate system – is one or two additional iteration steps, see Table 2. On the other hand, the simplicity of the proposed approach minimizes the risk of hidden bugs in the software. In the era of high-performance computers, there is no reason not to use this model in all kinds of scientific, engineering, and cadastral applications.

The main advantages of the proposed rigorous functional model of adjustment of horizontal geodetic networks in a projected coordinate system can be summarized as follows: highest accuracy, simplicity, and universality. This approach could lead to some other innovative solutions in geodesy, surveying, navigation, and positioning based on measured distances and directions or azimuths.

# Supplementary material to
# “Rigorous Computation of Classical Horizontal Geodetic Networks”

Sandi BERK[1] and Bojan STOPAR[2]

This supplement contains source code of a self-developed computer program used for testing the proposed models of adjustment of classical horizontal geodetic networks. A simple console application is written in standard C++ programming language. Double-precision floating-point arithmetic is used.

The application project consists of four files:

- AdNet.cpp
- Stdafx.h
- MTRX.hpp
- MTRX.cpp

The main source code file **AdNet.cpp** is given on pages S-2 to S-44. This file should be compiled together with the automatically generated **Stdafx.h**, and a self-developed library for matrix arithmetic. Its header file **MTRX.hpp** with class declarations is given on page S-45. The library source code file **MTRX.cpp** that contains the implementation of the matrix arithmetic algorithms is not given here. These algorithms can be taken from Press et al. (1992), for example.

The program enables processing of the test network (see page 16 in the paper: Fictitious Test Network) with the error-free and error-prone observation sets (see page 17 in the paper: Simulation of Measurement Campaigns). The network adjustment can be performed in the geodetic coordinate system and in three projected coordinate systems (see pages 14 and 15 in the paper: Map projections Used for Testing).

In case of the network adjustment in the geodetic coordinate system, the final coordinates and standard confidence ellipses of the new network points are rigorously transformed to all three projected coordinate systems. By running the program with different triplets of input parameters (see command-line arguments explanation on page S-19 of this Supplement), all processing results can be generated that are presented in Tables 1 to 8 in the paper.

[1] Surveying and Mapping Authority of the Republic of Slovenia, Zemljemerska ulica 12, SI-1000 Ljubljana, Slovenia ORCID: https://orcid.org/0000-0002-5074-6738 (corresponding author), email: sandi.berk@gov.si
[2] Prof., University of Ljubljana, Faculty of Civil and Geodetic Engineering, Jamova cesta 2, SI-1000 Ljubljana, Slovenia ORCID: https://orcid.org/0000-0003-3119-9967

```cpp
// AdNet.cpp (defines the entry point for the console application)
// ********************************************************************************************************
// RIGOROUS COMPUTATION OF CLASSICAL HORIZONTAL GEODETIC NETWORKS
// horizontal geodetic network adjustment in the geodetic coordinate system (longitude, latitude) or in a projected coordinate system (easting, northing)
//  - based on the Conformal Cylindrical, Equal-Area Cylindrical & Transverse Mercator projections
//  - using examples of error-free or error-prone (simulated) observations from the test network (Fig. 2) in the paper


#include "Stdafx.h"
#include "MTRX.hpp"

#define SP(x) setprecision(x)
#define SW(x) setw(x)

using namespace std;

// 'full' is used as a synonym for 'long double' data type (defined in MTRX.hpp, see the last page)

const full pi = 3.14159265358979323846L;                        // Archimedes’ constant
const full a = 6378137.0L;                                      // major semi-axis of the GRS80 ellipsoid (exact) [m]
const full e2 = 6.69438002290e-3L;                              // squared first eccentricity of the GRS80 ellipsoid
const full b = a*sqrtl(1-e2);                                   // minor semi-axis of the GRS80 ellipsoid [m]
const full n = (a-b)/(a+b);                                     // third eccentricity of the GRS80 ellipsoid
const full rR = a/(1+n)*(1+n*n*(1+n*n*(1+n*n*(1+25*n*n/64)/4)/16)/4);   // rectifying radius of the GRS80 ellipsoid [m]
const full w = 1 + 1e-14L;                                      // 1.00000000000001 (a bit more than 1)

full al[9], bt[9];                                              // Karney series (alphas & betas) for the Transverse Mercator (TM) projection

// ********************************************************************************************************

struct network_point
{
    char N[20];                                   // network point label (number/name)
    full l;                                       // geodetic longitude ... λ [rad]
    full f;                                       // geodetic latitude  ... φ [rad]
    full h;                                       // ellipsoidal height ... h [m]
    network_point(void);                          // default network point constructor
    network_point(full, full, full, const char*); // network point constructor from the geodetic coordinates
};
```

```
struct observation
{
    int I;                                          // index of the first network baseline point (standpoint)
    int J;                                          // index of the second network baseline point (forepoint)
    full d;                                         // error-free observation - distance [m] or direction [rad]
    full D;                                         // error-prone observation - distance [m] or direction [rad]
    observation(void);                              // default observation constructor
    observation(int, int);                          // observation constructor from the network point indices
};

struct confidence_ellipse
{
    full a;                                         // major semi-axis of the network point standard confidence ellipse [m]
    full b;                                         // minor semi-axis of the network point standard confidence ellipse [m]
    full t;                                         // azimuth of the major semi-axis of the network point standard confidence ellipse [rad]
    confidence_ellipse(void);                       // default standard confidence ellipse constructor
    confidence_ellipse(full, full, full);           // standard confidence ellipse constructor from the network point variances and covariances
};

full arctan2g(full y, full x)
// azimuth from coordinate differences in the local geodetic system (Eq. 16) on the interval [0, 2π)
{
    if (y >= 0) return pi / 2 - 2 * atanl(x / (sqrtl(x * x + y * y) + y));
    else return 3 * pi / 2 + 2 * atanl(x / (sqrtl(x * x + y * y) - y));
}

full angle_between_0_and_2pi(full x)
// conversion of an angle to the interval [0, 2π)
{
    return x - 2 * pi * (long(x / 2 / pi) - (x < 0));
}

inline full deg(full x)
// conversion from quasi-sexagesimal degrees into decimal degrees
{
    full y = ab(x);
    return sgn(x) * (250 * y - 60 * long(w * y) - long(100 * w * y)) / 90;
}

inline full dms(full x)
// conversion from decimal degrees into quasi-sexagesimal degrees
{
    full y = ab(x);
    return sgn(x) * (90 * y + 100 * long(w * y) + long(60 * w * y)) / 250;
}
```

```
network_point::network_point(void)
// default network point constructor
{
    strcpy_s(N, 18, ""), l = f = h = 0.0;
    return;
}

network_point::network_point(full la, full fi, full he, const char* na)
// network point constructor from the geodetic coordinates (la, fi, he)
{
    strcpy_s(N, 18, na), l = la, f = fi, h = he;
    return;
}

observation::observation(void)
// default observation constructor
{
    I = J = 0, d = D = 0.0;
    return;
}

observation::observation(int i, int j)
// observation constructor from the network point indices
{
    I = i, J = j, d = D = 0.0;
    return;
}

confidence_ellipse::confidence_ellipse(void)
// default standard confidence ellipse constructor
{
    a = b = t = 0.0;
    return;
}

confidence_ellipse::confidence_ellipse(full ee, full nn, full en)
// standard confidence ellipse constructor from the network point variance-covariance matrix elements (Eqs. 50–53)
// (Kuang 1996, pp. 164–165)
{
    full q = sqrtl(sq(ee - nn) + 4 * en * en);
    a = sqrtl((ee + nn + q) / 2), b = sqrtl((ee + nn - q) / 2);
    if (ee == nn && en == 0) t = 0.0;
    else t = arctan2g(2 * en, nn - ee) / 2;           // azimuth of the major semi-axis on the interval [0, π)
    return;
}
```

```
inline full Mc(full f)
// meridian radius of curvature at the given latitude (Eq. 2)
{
    return a * (1 - e2) / powl(1 - e2 * sq(sinl(f)), 1.5);
}

inline full Nc(full f)
// prime vertical radius of curvature at the given latitude (Eq. 3)
{
    return a / sqrtl(1 - e2 * sq(sinl(f)));
}

inline full Nc_dv_f(full f)
// derivative of the prime vertical radius of curvature with respect to the latitude (Eq. 4)
{
    return e2 * Nc(f) * sinl(f) * cosl(f) / (1 - e2 * sq(sinl(f)));
}

full psi(full &f)
// isometric latitude (Eq. 93)
{
    return asinhl(tanl(f)) - sqrtl(e2) * atanhl(sqrtl(e2) * sinl(f));
}

full qu(full &f)
// auxiliary parameter used to define the authalic latitude (Eq. 99)
{
    return (1 - e2) * (sinl(f) / (1 - e2 * sq(sinl(f))) + atanhl(sqrtl(e2) * sinl(f)) / sqrtl(e2));
}

full grid_distance_IJ(full ei, full ni, full ej, full nj)
// grid distance between the i-th and j-th network point (Eq. 60)
{
    return sqrtl(sq(ej - ei) + sq(nj - ni));
}

full grid_azimuth_IJ(full ei, full ni, full ej, full nj)
// grid azimuth from the i-th network point to the j-th network point (Eq. 61)
{
    return arctan2g(ej - ei, nj - ni);
}
```

```
// *****************************************************************************************************
// 3D Cartesian coordinates (X, Y, Z) from the geodetic coordinates (λ, φ, h) (Torge 2001, p. 100)

full X(network_point &A)
// Cartesian X coordinate of a network point from its geodetic coordinates (Eq. 116)
{
    return (Nc(A.f) + A.h) * cosl(A.f) * cosl(A.l);
}

full Y(network_point &A)
// Cartesian Y coordinate of a network point from its geodetic coordinates (Eq. 117)
{
    return (Nc(A.f) + A.h) * cosl(A.f) * sinl(A.l);
}

full Z(network_point &A)
// Cartesian Z coordinate of a network point from its geodetic coordinates (Eq. 118)
{
    return ((1 - e2) * Nc(A.f) + A.h) * sinl(A.f);
}

// *****************************************************************************************************
// Mapping equations for the Conformal Cylindrical (CC) projection (adapted to the network area with the centroid C)
// (Safari & Ardalan 2007; there is a typo on p. 140, Eq. 45 ... easting coordinate!)

full eCC(network_point &A, network_point &C)
// easting coordinate of a network point in the Conformal Cylindrical projection (adapted to the network area) from its geodetic coordinates (Eq. 91)
// C.l = longitude of natural origin (standard longitude), C.f = latitude of natural origin (standard latitude); C is the network centroid
{
    return Nc(C.f) * cosl(C.f) * (A.l - C.l);
}

full nCC(network_point &A, network_point &C)
// northing coordinate of a network point in the Conformal Cylindrical projection (adapted to the network area) from its geodetic coordinates (Eq. 92)
// C.l = longitude of natural origin (standard longitude), C.f = latitude of natural origin (standard latitude); C is the network centroid
{
    return Nc(C.f) * cosl(C.f) * (psi(A.f) - psi(C.f));
}
```

```
// *******************************************************************************************************************************************************
// Partial derivatives of the mapping equations for the Conformal Cylindrical (CC) projection (adapted to the network area): Jacobian matrix (Eq. 94)
// (Safari & Ardalan 2007)

full eCC_dv_l(network_point &A, network_point &C)
// partial derivative of the easting coordinate of a network point with respect to its longitude
{
    return Nc(C.f) * cosl(C.f);
}

// partial derivative of the easting coordinate of a network point with respect to its latitude is zero
// partial derivative of the northing coordinate of a network point with respect to its longitude is zero

full nCC_dv_f(network_point &A, network_point &C)
// partial derivative of the northing coordinate of a network point with respect to its latitude
{
    return Mc(A.f) / Nc(A.f) * Nc(C.f) * cosl(C.f) / cosl(A.f);
}

// *******************************************************************************************************************************************************
// Scale factor and meridian (grid) convergence for the Conformal Cylindrical (CC) projection (adapted to the network area)
// (Safari & Ardalan 2007)

full CC_scale(network_point &A, network_point &C)
// scale factor for a network point (Eq. 95)
// C.l = longitude of natural origin (standard longitude), C.f = latitude of natural origin (standard latitude); C is the network centroid
{
    return Nc(C.f) / Nc(A.f) * cosl(C.f) / cosl(A.f);
}

// meridian (grid) convergence for a network point is zero (Eq. 96)
```

```
// ***********************************************************************************************************
// Inverse mapping equations for the Conformal Cylindrical (CC) projection adapted to the network area with the centroid C (Eqs. 103–105)
// (Bildirici 2017)

network_point lf_from_enCC(full E, full N, network_point &A, network_point &C)
// geodetic coordinates from the projected coordinates of a network point in the Conformal Cylindrical projection (adapted to the network area)
// A gives ellipsoidal height and approximate geodetic coordinates of the network point; the latter are needed for the first iteration ...
// C.l = longitude of natural origin (standard longitude), C.f = latitude of natural origin (standard latitude); C is the network centroid
{
    network_point P, Q = A;
    full det;
    int i = 0;
    do
    {
        ++i, P = Q;
        det = - eCC_dv_l(P, C) * nCC_dv_f(P, C);                                                    // Jacobian determinant
        Q.l = P.l + (eCC(P, C) - E) * nCC_dv_f(P, C) / det;
        Q.f = P.f + (nCC(P, C) - N) * eCC_dv_l(P, C) / det;
    }
    while (ab(Q.l - P.l) + ab(Q.f - P.f) > 1e-15L && i < 200);
    return Q;
}

// ***********************************************************************************************************
// Mapping equations for the Equal-Area Cylindrical (EAC) projection (adapted to the network area with the centroid C)
// (Safari & Ardalan 2007) ... equation for the northing coordinate below is a bit modified; auxiliary parameter (qu) is applied, which is used to define
// the authalic latitude (Snyder 1987, p. 101)

full eEAC(network_point &A, network_point &C)
// easting coordinate of a network point in the Equal-Area Cylindrical projection (adapted to the network area) from its geodetic coordinates (Eq. 97)
// C.l = longitude of natural origin (standard longitude), C.f = latitude of natural origin (standard latitude); C is the network centroid
{
    return Nc(C.f) * cosl(C.f) * (A.l - C.l);
}

full nEAC(network_point &A, network_point &C)
// northing coordinate of a network point in the Equal-Area Cylindrical projection (adapted to the network area) from its geodetic coordinates (Eq. 98)
// C.l = longitude of natural origin (standard longitude), C.f = latitude of natural origin (standard latitude); C is the network centroid
{
    return sq(a) * (qu(A.f) - qu(C.f)) / Nc(C.f) / cosl(C.f) / 2;
}
```

```
// ********************************************************************************************************
// Partial derivatives of the mapping equations for the Equal-Area Cylindrical (EAC) projection (adapted to the network area): Jacobian matrix (Eq. 100)
// (Safari & Ardalan 2007; there is a typo on p. 138, Eq. 27 ... partial derivative of the northing coordinate!)

full eEAC_dv_l(network_point &A, network_point &C)
// partial derivative of the easting coordinate of a network point with respect to its longitude
{
    return Nc(C.f) * cosl(C.f);
}

// partial derivative of the easting coordinate of a network point with respect to its latitude is zero
// partial derivative of the northing coordinate of a network point with respect to its longitude is zero

full nEAC_dv_f(network_point &A, network_point &C)
// partial derivative of the northing coordinate of a network point with respect to its latitude
{
    return Mc(A.f) / Nc(C.f) * Nc(A.f) * cosl(A.f) / cosl(C.f);
}

// ********************************************************************************************************
// Inverse mapping equations for the Equal-Area Cylindrical (EAC) projection adapted to the network area with the centroid C (Eqs. 103–105)
// (Bildirici 2017)

network_point lf_from_enEAC(full E, full N, network_point &A, network_point &C)
// geodetic coordinates from the projected coordinates of a network point in the Equal-Area Cylindrical projection (adapted to the network area)
// A gives ellipsoidal height and approximate geodetic coordinates of the network point; the latter are needed for the first iteration ...
// C.l = longitude of natural origin (standard longitude), C.f = latitude of natural origin (standard latitude); C is the network centroid
{
    network_point P, Q = A;
    full det;
    int i = 0;
    do
    {
        ++i, P = Q;
        det = eEAC_dv_l(P, C) * nEAC_dv_f(P, C);                                    // Jacobian determinant
        Q.l = P.l - (eEAC(P, C) - E) * nEAC_dv_f(P, C) / det;
        Q.f = P.f - (nEAC(P, C) - N) * eEAC_dv_l(P, C) / det;
    }
    while (ab(Q.l - P.l) + ab(Q.f - P.f) > 1e-15L && i < 200);
    return Q;
}
```

```
// ***************************************************************************************************************
// Mapping equations for the Transverse Mercator (TM) projection (Karney 2011)

full eTM(network_point &A, full l0, full k0, full fe)
// easting coordinate of a network point in the Transverse Mercator projection (with the selected parameters) from its geodetic coordinates
// l0 = latitude of natural origin (central meridian), k0 = scale factor at natural origin, fe = false easting
{
    full tau = tanl(A.f);
    full sig = sinhl(sqrtl(e2) * atanhl(sqrtl(e2) * tau / sqrtl(1 + tau * tau)));
    full tau_ = tau * sqrtl(1 + sig * sig) - sig * sqrtl(1 + tau * tau);
    full ksi_ = atanl(tau_ / cosl(A.l - l0));
    full eta_ = asinhl(sinl(A.l - l0) / sqrtl(tau_ * tau_ + sq(cosl(A.l - l0))));
    full eta = eta_;
    for (int i = 1; i <= 8; i++) eta += ::al[i] * cosl(2 * i * ksi_) * sinhl(2 * i * eta_);
    return k0 * rR * eta + fe;
}

full nTM(network_point &A, full l0, full k0, full fn)
// northing coordinate of a network point in the Transverse Mercator projection (with the selected parameters) from its geodetic coordinates
// l0 = latitude of natural origin (central meridian), k0 = scale factor at natural origin, fn = false northing
{
    full tau = tanl(A.f);
    full sig = sinhl(sqrtl(e2) * atanhl(sqrtl(e2) * tau / sqrtl(1 + tau * tau)));
    full tau_ = tau * sqrtl(1 + sig * sig) - sig * sqrtl(1 + tau * tau);
    full ksi_ = atanl(tau_ / cosl(A.l - l0));
    full eta_ = asinhl(sinl(A.l - l0) / sqrtl(tau_ * tau_ + sq(cosl(A.l - l0))));
    full ksi = ksi_;
    for (int i = 1; i <= 8; i++) ksi += ::al[i] * sinl(2 * i * ksi_) * coshl(2 * i * eta_);
    return k0 * rR * ksi + fn;
}
```

```
// ****************************************************************************************************************************
// Scale factor and meridian (grid) convergence for the Transverse Mercator (TM) projection (with the selected parameters)
// (Kawase 2013)

full TM_scale(network_point &A, full l0, full k0)
// scale factor for a network point
// l0 = latitude of natural origin (central meridian), k0 = scale factor at natural origin
{
    int i;
    full sig = 0, tau = 0;
    full s = 2 * sqrtl(n) / (1 + n);
    full t = sinhl(atanhl(sinl(A.f)) - s * atanhl(s * sinl(A.f)));
    full ksi_ = atanl(t / cosl(A.l - l0));
    full eta_ = atanhl(sinl(A.l - l0) / sqrtl(1 + t * t));
    for (i = 1; i <= 8; i++) sig += ::al[i] * i * cosl(2 * i * ksi_) * coshl(2 * i * eta_);
    for (i = 1; i <= 8; i++) tau += ::al[i] * i * sinl(2 * i * ksi_) * sinhl(2 * i * eta_);
    return k0 * rR / a * sqrtl((1 + sq(tanl(A.f) * (1 - n) / (1 + n))) * (sq(1 + 2 * sig) + sq(2 * tau)) / (sq(t) + sq(cosl(A.l - l0))));
}

full TM_converg(network_point &A, full l0)
// meridian (grid) convergence for a network point
// l0 = latitude of natural origin (central meridian)
{
    int i;
    full sig = 0, tau = 0;
    full s = 2 * sqrtl(n) / (1 + n);
    full t = sinhl(atanhl(sinl(A.f)) - s * atanhl(s * sinl(A.f)));
    full ksi_ = atanl(t / cosl(A.l - l0));
    full eta_ = atanhl(sinl(A.l - l0) / sqrtl(1 + t * t));
    for (i = 1; i <= 8; i++) sig += ::al[i] * i * cosl(2 * i * ksi_) * coshl(2 * i * eta_);
    for (i = 1; i <= 8; i++) tau += ::al[i] * i * sinl(2 * i * ksi_) * sinhl(2 * i * eta_);
    return atanl((2 * tau * sqrtl(1 + t * t) + t * (1 + 2 * sig) * tanl(A.l - l0)) / ((1 + 2 * sig) * sqrtl(1 + t * t) - 2 * t * tau * tanl(A.l - l0)));
}
```

```
// ******************************************************************************************************************************
// Inverse mapping equations for the Transverse Mercator (TM) projection (Karney 2011)

network_point lf_from_enTM(full E, full N, network_point &A, full l0, full k0, full fe, full fn)
// geodetic coordinates from the projected coordinates of a network point in the Transverse Mercator projection (with the selected parameters)
// A gives ellipsoidal height of the network point
// l0 = latitude of natural origin (central meridian), k0 = scale factor at natural origin, fe = false easting, fn = false northing
{
    int i;
    network_point Q = A;
    full eta = (E - fe) / k0 / rR, ksi = (N - fn) / k0 / rR;
    full eta_ = eta;
    for (i = 1; i <= 8; i++) eta_ -= ::bt[i] * cosl(2 * i * ksi) * sinhl (2 * i * eta);
    full ksi_ = ksi;
    for (i = 1; i <= 8; i++) ksi_ -= ::bt[i] * sinl(2 * i * ksi) * coshl (2 * i * eta);
    full tau_ = sinl(ksi_) / sqrtl(sq(sinhl(eta_)) + sq(cosl(ksi_)));
    full sig, tau = tau_;
    // Newton-Raphson method:
    i = 0;
    do
    {
        sig = sinhl(sqrtl(e2) * atanh(sqrtl(e2) * tau / sqrtl(1 + tau * tau)));
        tau -= (tau * sqrtl(1 + sig * sig) - sig * sqrtl(1 + tau * tau) - tau_) / ((sqrtl((1 + sig * sig) * (1 + tau * tau)) - sig * tau) * (1 - e2)
            * sqrtl(1 + tau * tau) / (1 + (1 - e2) * tau * tau));
    }
    while (++i < 5);              // usually, two iterations are sufficient here
    Q.l = atanl(sinhl(eta_) / cosl(ksi_)) + l0;
    Q.f = atanl(tau);
    return Q;
}

// ******************************************************************************************************************************

full al_IJ(network_point &I, network_point &J)
// auxiliary parameter for the network side IJ (Eq. 9)
{
    return cosl(I.f) * sinl(J.f) - sinl(I.f) * cosl(J.f) * cosl(J.l - I.l);
}

full be_IJ(network_point &I, network_point &J)
// auxiliary parameter for the network side IJ (Eq. 10)
{
    return sinl(I.f) * sinl(J.f) + cosl(I.f) * cosl(J.f) * cosl(J.l - I.l);
}
```

```
full deX_IJ(network_point &I, network_point &J)
// X coordinate difference in the local geodetic system for the network side IJ (Eq. 11)
{
    return al_IJ(I, J) * (Nc(J.f) + J.h) - e2 * (Nc(J.f) * sinl(J.f) - Nc(I.f) * sinl(I.f)) * cosl(I.f);
}

full deY_IJ(network_point &I, network_point &J)
// Y coordinate difference in the local geodetic system for the network side IJ (Eq. 12)
{
    return (Nc(J.f) + J.h) * cosl(J.f) * sinl(J.l - I.l);
}

full deZ_IJ(network_point &I, network_point &J)
// Z coordinate difference in the local geodetic system for the network side IJ (Eq. 13)
{
    return be_IJ(I, J) * (Nc(J.f) + J.h) - e2 * (Nc(J.f) * sinl(J.f) - Nc(I.f) * sinl(I.f)) * sinl(I.f) - Nc(I.f) - I.h;
}
full distance_IJ(network_point &I, network_point &J)
// spatial straight-line distance between the network points I and J (Eq. 14)
{
 // return sqrtl(sq(deX_IJ(I, J)) + sq(deY_IJ(I, J)) + sq(deZ_IJ(I, J)));
    return sqrtl(sq(X(J) - X(I)) + sq(Y(J) - Y(I)) + sq(Z(J) - Z(I)));        // preferable alternative!
}

full azimuth_IJ(network_point &I, network_point &J)
// azimuth in the local geodetic system from the network point I to the network point J (Eq. 15)
{
    return arctan2g(deY_IJ(I, J), deX_IJ(I, J));
}

full ro_IJ(network_point &I, network_point &J)
// rho - an auxiliary parameter for the network side IJ (Eq. 20)
{
    return deX_IJ(I, J) / (sq(deX_IJ(I, J)) + sq(deY_IJ(I, J)));
}

full si_IJ(network_point &I, network_point &J)
// sigma - an auxiliary parameter for the network side IJ (Eq. 21)
{
    return deY_IJ(I, J) / (sq(deX_IJ(I, J)) + sq(deY_IJ(I, J)));
}
```

```
full ch_IJ(network_point &I, network_point &J)
// chi - an auxiliary parameter for the network side IJ (Eq. 22)
{
    return deX_IJ(I, J) / distance_IJ(I, J);
}

full ps_IJ(network_point &I, network_point &J)
// psi - an auxiliary parameter for the network side IJ (Eq. 23)
{
    return deY_IJ(I, J) / distance_IJ(I, J);
}

full xi_IJ(network_point &I, network_point &J)
// xi - an auxiliary parameter for the network side IJ (Eq. 24)
{
    return deZ_IJ(I, J) / distance_IJ(I, J);
}

full ep_IJ(network_point &I, network_point &J)
// epsilon - an auxiliary quantity for the network side IJ (Eq. 25)
{
    return sinl(I.f) * cosl(J.f) - cosl(I.f) * sinl(J.f) * cosl(J.l - I.l);
}

full ze_IJ(network_point &I, network_point &J)
// zeta - an auxiliary parameter for the network side IJ (Eq. 26)
{
    return cosl(I.f) * cosl(J.f) + sinl(I.f) * sinl(J.f) * cosl(J.l - I.l);
}

full et_IJ(network_point &I, network_point &J)
// eta - an auxiliary parameter for the network side IJ (Eq. 27)
{
    return e2 * (Nc(I.f) * cosl(2 * I.f) + (Nc(J.f) * sinl(J.f) + Nc_dv_f(I.f) * cosl(I.f)) * sinl(I.f));
}

full th_IJ(network_point &I, network_point &J)
// theta - an auxiliary parameter for the network side IJ (Eq. 28)
{
    return (Nc_dv_f(J.f) * cosl(J.f) - (Nc(J.f) + J.h) * sinl(J.f)) * sinl(J.l - I.l);
}
```

```
full my_IJ(network_point &I, network_point &J)
// my - an auxiliary parameter for the network side IJ (Eq. 29)
{
    return Nc(J.f) * cosl(J.f) + Nc_dv_f(J.f) * sinl(J.f);
}

full ka_IJ(network_point &I, network_point &J)
// kappa - an auxiliary parameter for the network side IJ (Eq. 30)
{
    return ze_IJ(I, J) * (Nc(J.f) + J.h) + Nc_dv_f(J.f) * al_IJ(I, J) - e2 * my_IJ(I, J) * cosl(I.f);
}

// ************************************************************************************************************
// Partial derivatives of the distance and direction observables for the network adjustment in a geodetic coordinate system (longitude, latitude)

full Dis_dv_Il(network_point &I, network_point &J)
// partial derivative of the distance observable with respect to the standpoint longitude (Eq. 31)
{
    return (xi_IJ(I, J) * cosl(I.f) - ch_IJ(I, J) * sinl(I.f)) * deY_IJ(I, J) - ps_IJ(I, J) * (Nc(J.f) + J.h) * cosl(J.f) * cosl(J.l - I.l);
}

full Dis_dv_If(network_point &I, network_point &J)
// partial derivative of the distance observable with respect to the standpoint latitude (Eq. 32)
{
    return xi_IJ(I, J) * deX_IJ(I, J) - ch_IJ(I, J) * (be_IJ(I, J) * (Nc(J.f) + J.h) - et_IJ(I, J));
}

full Dis_dv_Jl(network_point &I, network_point &J)
// partial derivative of the distance observable with respect to the forepoint longitude (Eq. 33)
{
    return -Dis_dv_Il(I, J);
}

full Dis_dv_Jf(network_point &I, network_point &J)
// partial derivative of the distance observable with respect to the forepoint latitude (Eq. 34)
{
    return xi_IJ(I, J) * (ep_IJ(I, J) * (Nc(J.f) + J.h) + Nc_dv_f(J.f) * be_IJ(I, J) - e2 * my_IJ(I, J) * sinl(I.f)) + ps_IJ(I, J) * th_IJ(I, J) + ch_IJ(I, J)
        * ka_IJ(I, J);
}
full Dir_dv_Il(network_point &I, network_point &J)
// partial derivative of the direction observable with respect to the standpoint longitude (Eq. 35)
{
    return si_IJ(I, J) * sinl(I.f) * deY_IJ(I, J) - ro_IJ(I, J) * (Nc(J.f) + J.h) * cosl(J.f) * cosl(J.l - I.l);
}
```

```
full Dir_dv_If(network_point &I, network_point &J)
// partial derivative of the direction observable with respect to the standpoint latitude (Eq. 36)
{
    return si_IJ(I, J) * (be_IJ(I, J) * (Nc(J.f) + J.h) - et_IJ(I, J));
}

full Dir_dv_Jl(network_point &I, network_point &J)
// partial derivative of the direction observable with respect to the forepoint longitude (Eq. 37)
{
    return -Dir_dv_Il(I, J);
}

full Dir_dv_Jf(network_point &I, network_point &J)
// partial derivative of the direction observable with respect to the forepoint latitude (Eq. 38)
{
    return ro_IJ(I, J) * th_IJ(I, J) - si_IJ(I, J) * ka_IJ(I, J);
}

// *****************************************************************************************************************
// Partial derivatives of the distance and direction observables for the network adjustment in the projected coordinate system (easting, northing)

full Dis_dv_Ie(full ei, full ni, full ej, full nj)
// partial derivative of the distance observable with respect to the standpoint easting (Eq. 71)
{
    return -(ej - ei) / grid_distance_IJ(ei, ni, ej, nj);
}

full Dis_dv_In(full ei, full ni, full ej, full nj)
// partial derivative of the distance observable with respect to the standpoint northing (Eq. 72)
{
    return -(nj - ni) / grid_distance_IJ(ei, ni, ej, nj);
}

full Dis_dv_Je(full ei, full ni, full ej, full nj)
// partial derivative of the distance observable with respect to the forepoint easting (Eq. 73)
{
    return -Dis_dv_Ie(ei, ni, ej, nj);
}

full Dis_dv_Jn(full ei, full ni, full ej, full nj)
// partial derivative of the distance observable with respect to the forepoint northing (Eq. 74)
{
    return -Dis_dv_In(ei, ni, ej, nj);
}
```

```
full Dir_dv_Ie(full ei, full ni, full ej, full nj)
// partial derivative of the direction observable with respect to the standpoint easting (Eq. 75)
{
    return -(nj - ni) / (sq(ej - ei) + sq(nj - ni));
}

full Dir_dv_In(full ei, full ni, full ej, full nj)
// partial derivative of the direction observable with respect to the standpoint northing (Eq. 76)
{
    return (ej - ei) / (sq(ej - ei) + sq(nj - ni));
}

full Dir_dv_Je(full ei, full ni, full ej, full nj)
// partial derivative of the direction observable with respect to the forepoint easting (Eq. 77)
{
    return -Dir_dv_Ie(ei, ni, ej, nj);
}

full Dir_dv_Jn(full ei, full ni, full ej, full nj)
// partial derivative of the direction observable with respect to the forepoint northing (Eq. 78)
{
    return -Dir_dv_In(ei, ni, ej, nj);
}

// ************************************************************************************************************************************************************
```

```
// ***********************************************************************************************************
// ***********************************************************************************************************
// ***********************************************************************************************************

int main(int argc, char* argv[])
{
    // Karney series (alphas & betas) for the Transverse Mercator (TM) projection (Karney 2011):

    al[1] = n/2-2*n*n/3+5*powl(n,3)/16+41*powl(n,4)/180-127*powl(n,5)/288+7891*powl(n,6)/37800+72161*powl(n,7)/387072-18975107*powl(n,8)/50803200;
    al[2] = 13*n*n/48-3*powl(n,3)/5+557*powl(n,4)/1440+281*powl(n,5)/630-1983433*powl(n,6)/1935360+13769*powl(n,7)/28800+148003883*powl(n,8)/174182400;
    al[3] = 61*powl(n,3)/240-103*powl(n,4)/140+15061*powl(n,5)/26880+167603*powl(n,6)/181440-67102379*powl(n,7)/29030400+79682431*powl(n,8)/79833600;
    al[4] = 49561*powl(n,4)/161280-179*powl(n,5)/168+6601661*powl(n,6)/7257600+97445*powl(n,7)/49896-40176129013.*powl(n,8)/7664025600.;
    al[5] = 34729*powl(n,5)/80640-3418889*powl(n,6)/1995840+14644087*powl(n,7)/9123840+2605413599.*powl(n,8)/622702080;
    al[6] = 212378941.*powl(n,6)/319334400-30705481*powl(n,7)/10378368+175214326799.*powl(n,8)/58118860800.;
    al[7] = 1522256789.*powl(n,7)/1383782400.-16759934899.*powl(n,8)/3113510400.;
    al[8] = 1424729850961.*powl(n,8)/743921418240.;

    bt[1] = n/2-2*n*n/3+37*powl(n,3)/96-powl(n,4)/360-81*powl(n,5)/512+96199*powl(n,6)/604800-5406467*powl(n,7)/38707200+7944359*powl(n,8)/67737600;
    bt[2] = n*n/48+powl(n,3)/15-437*powl(n,4)/1440+46*powl(n,5)/105-1118711*powl(n,6)/3870720+51841*powl(n,7)/1209600+24749483*powl(n,8)/348364800;
    bt[3] = 17*powl(n,3)/480-37*powl(n,4)/840-209*powl(n,5)/4480+5569*powl(n,6)/90720+9261899*powl(n,7)/58060800-6457463*powl(n,8)/17740800;
    bt[4] = 4397*powl(n,4)/161280-11*powl(n,5)/504-830251*powl(n,6)/7257600+466511*powl(n,7)/2494800+324154477*powl(n,8)/7664025600.;
    bt[5] = 4583*powl(n,5)/161280-108847*powl(n,6)/3991680-8005831*powl(n,7)/63866880+22894433*powl(n,8)/124540416;
    bt[6] = 20648693*powl(n,6)/638668800-16363163*powl(n,7)/518918400-2204645983.*powl(n,8)/12915302400.;
    bt[7] = 219941297*powl(n,7)/5535129600.-497323811*powl(n,8)/12454041600.;
    bt[8] = 191773887257.*powl(n,8)/3719607091200.;

    cout.setf(ios::right | ios::showpoint | ios::fixed);

    if (argc != 4)
    {
        // ERROR Message for an invalid program call
        cout << endl << " AdNet ERROR: Invalid program call!" << endl;
        cout << " Example:   ...path>.\\AdNet 1 2 3" << endl;
        cout << "   1st parameter:" << endl;
        cout << "     1 ... geodetic coordinate system (longitude, latitude)" << endl;
        cout << "     2 ... projected coordinate system (easting, northing): Conformal Cylindrical projection" << endl;
        cout << "     3 ... projected coordinate system (easting, northing): Equal-Area Cylindrical projection" << endl;
        cout << "     4 ... projected coordinate system (easting, northing): Transverse Mercator projection" << endl;
        cout << "   2nd parameter:" << endl;
        cout << "     1 ... error-free observations" << endl;
        cout << "     2 ... error-prone (simulated) observations" << endl;
        cout << "   3rd parameter:" << endl;
        cout << "     n ... number of iterations [1..100]" << endl << endl;
        return 0;
    }
```

```
int cs;     // coordinate system (1 = geodetic, 2 = CC projection, 3 = EAC projection, 4 = TM projection)
int ob;     // observation errors (1 = error-free observations, 2 = error-prone observations)
int it;     // number of adjustment iterations to be done [1..100]

cs = atoi(argv[1]);
if (cs < 1 || cs > 4)
{
    // ERROR Message for invalid 1st parameter
    cout << endl << " AdNet ERROR: Invalid 1st parameter (coordinate system)!" << endl;
    cout << " Available options are:" << endl;
    cout << "     1 ... geodetic coordinate system (longitude, latitude)" << endl;
    cout << "     2 ... projected coordinate system (easting, northing): Conformal Cylindrical projection" << endl;
    cout << "     3 ... projected coordinate system (easting, northing): Equal-Area Cylindrical projection" << endl;
    cout << "     4 ... projected coordinate system (easting, northing): Transverse Mercator projection" << endl;
    cout << endl;
    return 0;
}

ob = atoi(argv[2]);
if (ob < 1 || ob > 2)
{
    // ERROR Message for invalid 2nd parameter
    cout << endl << " AdNet ERROR: Invalid 2nd parameter (observation errors)!" << endl;
    cout << " Available options are:" << endl;
    cout << "     1 ... error-free observations" << endl;
    cout << "     2 ... error-prone (simulated) observations" << endl;
    cout << endl;
    return 0;
}

it = atoi(argv[3]);
if (it < 1 || it > 100)
{
    // ERROR Message for invalid 3rd parameter
    cout << endl;
    cout << " AdNet ERROR: Invalid 3rd parameter (number of adjustment iterations)!" << endl;
    cout << " It should be an integer from 1 to 100." << endl;
    cout << endl;
    return 0;
}
```

```
// ********************************************************************************************************************************************************
// NETWORK PARAMETERS

const int n_dis = 9;                                   // number of distance observations
const int n_dir = 18;                                  // number of direction observations
const int n_fix = 2;                                   // number of fixed/known points
const int n_new = 4;                                   // number of new/unknown points

const int n_obs = n_dis + n_dir;                       // total number of observations
const int n_pts = n_fix + n_new;                       // total number of network points
const int n_unk = n_pts + 2 * n_new;                   // total number of unknowns
const int n_red = n_obs - n_unk;                       // number of redundant observations

network_point NP[n_pts + 1];                           // set of network points with exact geodetic coordinates
network_point aNP[n_new + 1];                          // set of new network points with approximate geodetic coordinates
network_point NC;                                      // network centroid with geodetic coordinates
network_point I, J;                                    // network points for the current network side (standpoint & forepoint)
observation O[n_obs + 1];                              // set of observations (topology only)
full OR[n_pts + 1];                                    // set of exact orientations
full sOR[n_pts + 1];                                   // set of simulated orientations
full eNP[n_pts + 1], nNP[n_pts + 1];                   // sets of projected coordinates of network points (from exact geodetic coordinates)
full ei, ni, ej, nj;                                   // projected coordinates of network points for the current network side (standpoint & forepoint)

full tmp;                                              // temporal value
long i, j;                                             // counters

// ********************************************************************************************************************************************************
// NETWORK GEOMETRY

// Exact geodetic coordinates of selected network points
// ... longitude and latitude are given in quasi-sexagesimal degrees (degrees + 2 digits for arcminutes + 2 digits for arcseconds)
// Points with numbers:
//   - 1 to 4 are new network points (unknown stations)
//   - 5 & 6 are fixed network points (known stations)

NP[1] = {  9.3314, 47.0855, 1934.0, "Alpspitz" };      // Alpspitz, Eastern Alps, Liechtenstein
NP[2] = { 13.5012, 46.2242, 2864.0, "Triglav" };       // Triglav, Julian Alps, Slovenia
NP[3] = { 11.5202, 46.1500, 3192.0, "Vezzana" };       // Vezzana, Dolomites, Italy
NP[4] = { 10.5907, 47.2516, 2962.0, "Zugspitze" };     // Zugspitze, Wetterstein Mountains, Germany
NP[5] = { 12.4143, 47.0430, 3798.0, "Grossglockner" }; // Grossglockner, High Tauern, Austria
NP[6] = { 10.0556, 46.2002, 2862.0, "Sassalb" };       // Sassalb, Livigno Alps, Switzerland
```

```
// Parameters of projected coordinate systems:
NC = { 11.40, 46.50, 0.0, "" };                 // network centroid (longitude and latitude of natural origin for CC and EAC projections)
full L0 = 12 * pi / 180;                        // latitude of natural origin (central meridian) is 12 degrees (for TM projection)
full K0 = 0.9998;                               // scale factor at natural origin is 0.9998 (for TM projection)
full E0 = 500000.0;                             // false easting is 500000 m (for TM projection)
full N0 = -5000000.0;                           // false northing is -5000000 m (for TM projection)

//
*************************************************************************************************************************************************************************
// NETWORK TOPOLOGY

// Distance observations:
O[1] = { 1, 4 };    // Alpspitz-Zugspitze
O[2] = { 1, 6 };    // Alpspitz-Sassalb
O[3] = { 2, 3 };    // Triglav-Vezzana
O[4] = { 2, 5 };    // Triglav-Grossglockner
O[5] = { 3, 4 };    // Vezzana-Zugspitze
O[6] = { 3, 5 };    // Vezzana-Grossglockner
O[7] = { 3, 6 };    // Vezzana-Sassalb
O[8] = { 4, 5 };    // Zugspitze-Grossglockner
O[9] = { 4, 6 };    // Zugspitze-Sassalb

// Direction observations:
O[10] = { 1, 4 };   // Alpspitz-Zugspitze
O[11] = { 1, 6 };   // Alpspitz-Sassalb
O[12] = { 2, 3 };   // Triglav-Vezzana
O[13] = { 2, 5 };   // Triglav-Grossglockner
O[14] = { 3, 6 };   // Vezzana-Sassalb
O[15] = { 3, 4 };   // Vezzana-Zugspitze
O[16] = { 3, 5 };   // Vezzana-Grossglockner
O[17] = { 3, 2 };   // Vezzana-Triglav
O[18] = { 4, 5 };   // Zugspitze-Grossglockner
O[19] = { 4, 3 };   // Zugspitze-Vezzana
O[20] = { 4, 6 };   // Zugspitze-Sassalb
O[21] = { 4, 1 };   // Zugspitze-Alpspitz
O[22] = { 5, 2 };   // Grossglockner-Triglav
O[23] = { 5, 3 };   // Grossglockner-Vezzana
O[24] = { 5, 4 };   // Grossglockner-Zugspitze
O[25] = { 6, 1 };   // Sassalb-Alpspitz
O[26] = { 6, 4 };   // Sassalb-Zugspitze
O[27] = { 6, 3 };   // Sassalb-Vezzana
```

```
// ***********************************************************************************************************************
// Conversion of geodetic coordinates of selected network points from quasi-sexagesimal degrees into decimal degrees

for (i = 1; i <= n_pts; i++) NP[i].l = deg(NP[i].l), NP[i].f = deg(NP[i].f);

NC.l = deg(NC.l), NC.f = deg(NC.f);

// ***********************************************************************************************************************
// Conversion of geodetic coordinates of selected network points from decimal degrees into radians

for (i = 1; i <= n_pts; i++) NP[i].l *= pi / 180, NP[i].f *= pi / 180;

NC.l *= pi / 180, NC.f *= pi / 180;

// ***********************************************************************************************************************
// NETWORK ADJUSTMENT REPORT

// ... stream to the network adjustment report
ofstream to_txt("AdNet.txt", ios::in | ios::trunc);
to_txt.setf(ios::right | ios::showpoint | ios::fixed);

to_txt << "RIGOROUS COMPUTATION OF CLASSICAL HORIZONTAL GEODETIC NETWORKS" << endl;
to_txt << "Adjustment in the ";
if (cs == 1) to_txt << "Geodetic";
else to_txt << "Projected";
to_txt << " Coordinate System";
if (cs == 2) to_txt << ": Conformal Cylindrical Projection";
if (cs == 3) to_txt << ": Equal-Area Cylindrical Projection";
if (cs == 4) to_txt << ": Transverse Mercator Projection";
to_txt << endl;
if (ob == 1) to_txt << "Error-Free Observations";
if (ob == 2) to_txt << "Error-Prone (Simulated) Observations";
to_txt << endl;
to_txt << it << " Adjustment Iterations" << endl;
to_txt << "(Program call: .\\AdNet " << cs << " " << ob << " " << it << ")" << endl << endl;

to_txt << "Number of distance observations  = " << n_dis << endl;
to_txt << "Number of direction observations = " << n_dir << endl;
to_txt << "Total number of observations     = " << n_obs << endl;
to_txt << "-------------------------------------" << endl;
```

```
to_txt << "Number of known/fixed points     = " << n_fix << endl;
to_txt << "Number of unknown/new points      = " << n_new << endl;
to_txt << "Number of coordinate unknowns    = " << 2 * n_new << endl;
to_txt << "Number of orientation unknowns   = " << n_pts << endl;
to_txt << "Total number of unknowns         = " << n_unk << endl;
to_txt << "-------------------------------------" << endl;
to_txt << "Number of redundant observations = " << n_red << endl << endl;

to_txt << "Exact geodetic coordinates of network points:" << endl;
to_txt << " Pt                 Longitude    [deg, dms]                  Latitude    [deg, dms] Height [m]   Location        " << endl;
to_txt << "-----------------------------------------------------------------------------------------------------------------" << endl;
for (i = 1; i <= n_pts; i++) to_txt << SW(3) << i << SP(13) << SW(20) << 180 * NP[i].l / pi << SW(20) << dms(180 * NP[i].l / pi) << SW(20) << 180 * NP[i].f
    / pi << SW(20) << dms(180 * NP[i].f / pi) << SP(3) << SW(11) << NP[i].h << "   " << NP[i].N << endl;

if(cs > 1) to_txt << SW(3) << "C" << SP(13) << SW(20) << 180 * NC.l / pi << SW(20) << dms(180 * NC.l / pi) << SW(20) << 180 * NC.f / pi << SW(20)
    << dms(180 * NC.f / pi) << SP(3) << SW(11) << 0.0 << "   " << "network centroid" << endl;
to_txt << endl;

if (cs == 4) // Transverse Mercator projection
{
    to_txt << "Central meridian = " << SP(1) << 180 * L0 / pi << " deg" << endl;
    to_txt << "Scale factor     = " << SP(4) << K0 << endl;
    to_txt << "False easting    = " << SP(1) << E0 << " m" << endl;
    to_txt << "False northing   = " << SP(1) << N0 << " m" << endl << endl;
}

if(cs > 1)          // Projected Coordinate System
{
    to_txt << "Projected coordinates of network points:" << endl;
    to_txt << " Pt           Easting [m]          Northing [m]" << endl;
    to_txt << "-----------------------------------------------" << endl;
    for (i = 1; i <= n_pts; i++)
    {
        if (cs == 2) eNP[i] = eCC(NP[i], NC), nNP[i] = nCC(NP[i], NC);
        if (cs == 3) eNP[i] = eEAC(NP[i], NC), nNP[i] = nEAC(NP[i], NC);
        if (cs == 4) eNP[i] = eTM(NP[i], L0, K0, E0), nNP[i] = nTM(NP[i], L0, K0, N0);
        to_txt << SW(3) << i << SP(11) << SW(22) << eNP[i] << SW(22) << nNP[i] << endl;
    }
    to_txt << endl;
}
```

```
// ************************************************************************************************************************************
// Generating error-free observations

// Distances and azimuths (from exact coordinates):
for (i = 1; i <= n_dis; i++) O[i].d = distance_IJ(NP[O[i].I], NP[O[i].J]);
for (i = n_dis + 1; i <= n_obs; i++) O[i].d = azimuth_IJ(NP[O[i].I], NP[O[i].J]);

to_txt << "Error-free spatial distances:" << endl;
to_txt << "Pt1 Pt2          Distance [m]" << endl;
to_txt << "-----------------------------" << endl;
for (i = 1; i <= n_dis; i++) to_txt << SW(3) << O[i].I << SW(4) << O[i].J << SP(11) << SW(22) << O[i].d << endl;
to_txt << endl;

to_txt << "Error-free azimuths:" << endl;
to_txt << "Pt1 Pt2                   Azimuth    [deg, dms]" << endl;
to_txt << "-----------------------------------------------" << endl;
for (i = n_dis + 1; i <= n_obs; i++) to_txt << SW(3) << O[i].I << SW(4) << O[i].J << SP(13) << SW(20) << 180 * O[i].d / pi << SW(20) << dms(180 * O[i].d
    / pi) << endl;
to_txt << endl;

// Exact orientation angles from error-free azimuths (to the first forepoint at each standpoint):
for (i = 1; i <= n_pts; i++) for (j = n_dis + 1; j <= n_obs; j++) if (O[j].I == i)
{
    OR[i] = O[j].d;
    break;
}

to_txt << "Error-free orientation angles:" << endl;
to_txt << " Pt               Orientation    [deg, dms]" << endl;
to_txt << "-------------------------------------------" << endl;
for (i = 1; i <= n_pts; i++) to_txt << SW(3) << i << SP(13) << SW(20) << 180 * OR[i] / pi << SW(20) << dms(180 * OR[i] / pi) << endl;
to_txt << endl;

// Error-free reduced direction observations (direction to the first forepoint at each standpoint is set to 0):
for (i = 1; i <= n_pts; i++) for (j = n_dis + 1; j <= n_obs; j++) if (O[j].I == i) O[j].d = angle_between_0_and_2pi(O[j].d - OR[i]);

to_txt << "Error-free reduced directions:" << endl;
to_txt << "Pt1 Pt2                 Direction    [deg, dms]" << endl;
to_txt << "-----------------------------------------------" << endl;
for (i = n_dis + 1; i <= n_obs; i++) to_txt << SW(3) << O[i].I << SW(4) << O[i].J << SP(13) << SW(20) << 180 * O[i].d / pi << SW(20) << dms(180 * O[i].d
    / pi) << endl;
to_txt << endl;
```

```
matrix mA(n_obs, n_unk), vL(n_obs), mP(n_obs, n_obs), mAm(n_unk, n_unk), mH(n_unk, n_unk), vX0(n_unk), mQ, mQp, mQl, vdX, vX, vV;

// mA  ... network design matrix
// vL  ... misclosure vector (distances + directions)
// mP  ... weight matrix of observations
// mAm ... mapping design matrix
// mH  ... metric or Lamé matrix
// vX0 ... vector of approximate values of unknowns (coordinates + orientations)
// mQ  ... cofactor matrix of unknowns - original (i.e. in the coordinate system of the network adjustment)
// mQp ... cofactor matrix of unknovns transformed from the geodetic to the projected coordinate system (CC, EAC, or TM)
// mQl ... cofactor matrix of unknowns transformed from the geodetic to the local geodetic systems (for each new network point)
// vdX ... vector of corrections of unknowns
// vX  ... vector of estimated values of unknowns
// vV  ... vector of observation residuals

// ******************************************************************************************************
// Assigning (first) approximate values of the network unknowns (vX0)

// Approximate coordinates generated from exact coordinates in decimal degrees rounded to 2 digits (errors up to 18 arcseconds ... about 556 m):
for (i = 1; i <= n_new; i++)
{
    aNP[i] = NP[i];
    aNP[i].l = pi * round(18000 * NP[i].l / pi) / 18000;
    aNP[i].f = pi * round(18000 * NP[i].f / pi) / 18000;
}

// Approximate orientation angles generated from the calculated orientation angles in decimal degrees rounded to 2 digits (errors up to 18 arcseconds):
for (i = 2 * n_new + 1; i <= n_unk; i++) vX0(i) = pi * round(18000 * OR[i - 8] / pi) / 18000;

to_txt << "Approximate coordinates of new network points:" << endl;
if (cs == 1)        // Geodetic Coordinate System
{
    to_txt << " Pt                 Longitude    [deg, dms]                 Latitude    [deg, dms]" << endl;
    to_txt << "-----------------------------------------------------------------------------------" << endl;
    for (i = 1; i <= n_new; i++)
    {
        vX0(2 * i - 1) = aNP[i].l, vX0(2 * i) = aNP[i].f;
        to_txt << SW(3) << i << SP(13) << SW(20) << 180 * vX0(2 * i - 1) / pi << SW(20) << dms(180 * vX0(2 * i - 1) / pi) << SW(20) << 180 * vX0(2 * i)
            / pi << SW(20) << dms(180 * vX0(2 * i) / pi) << endl;
    }
}
```

```
else                // Projected Coordinate System
{
    to_txt << " Pt           Easting [m]          Northing [m]" << endl;
    to_txt << "-----------------------------------------------" << endl;
    for (i = 1; i <= n_new; i++)
    {
        if (cs == 2) vX0(2 * i - 1) = eCC(aNP[i], NC), vX0(2 * i) = nCC(aNP[i], NC);
        if (cs == 3) vX0(2 * i - 1) = eEAC(aNP[i], NC), vX0(2 * i) = nEAC(aNP[i], NC);
        if (cs == 4) vX0(2 * i - 1) = eTM(aNP[i], L0, K0, E0), vX0(2 * i) = nTM(aNP[i], L0, K0, N0);
        to_txt << SW(3) << i << SP(11) << SW(22) << vX0(2 * i - 1) << SW(22) << vX0(2 * i) << endl;
    }
}
to_txt << endl;
to_txt << "Approximate orientation angles:" << endl;
to_txt << " Pt               Orientation    [deg, dms]" << endl;
to_txt << "-------------------------------------------" << endl;
for (i = 1; i <= n_pts; i++) to_txt << SW(3) << i << SP(13) << SW(20) << 180 * vX0(2 * n_new + i) / pi << SW(20) << dms(180 * vX0(2 * n_new + i) / pi)
    << endl;
to_txt << endl;
vX = vX0; // assigning unknowns for the first iteration

// *****************************************************************************************************************
// Assigning error-free observations or introducing errors into error-free observations

if (ob == 1)      // error-free observations
{
    for (i = 1; i <= n_obs; i++) O[i].D = O[i].d;
}
else              // simulated observations
{
    // Distance generated from error-free distance rounded to nearest even decimeter (errors up to 10 centimeters, rms = 0.069 meters):
    for (i = 1; i <= n_dis; i++) O[i].D = round(5 * O[i].d) / 5;

    // Direction generated from error-free direction in decimal degrees rounded to 4 digits (errors up to 0.18 arcseconds, rms = 0.11 arcseconds):
    for (i = n_dis + 1; i <= n_obs; i++) O[i].D = pi * round(1800000 * O[i].d / pi) / 1800000;

    // Simulated orientation angles from simulated directions (to the first forepoint at each standpoint):
    for (i = 1; i <= n_pts; i++) for (j = n_dis + 1; j <= n_obs; j++) if (O[j].I == i)
    {
        sOR[i] = O[j].D;
        break;
    }
    // Simulated reduced direction observations (direction to the first forepoint at each standpoint is set to 0):
    for (i = 1; i <= n_pts; i++) for (j = n_dis + 1; j <= n_obs; j++) if (O[j].I == i) O[j].D = angle_between_0_and_2pi(O[j].D - sOR[i]);
}
```

```
to_txt << "Distance observations:" << endl;
to_txt << "Pt1 Pt2          Distance [m]" << endl;
to_txt << "-----------------------------" << endl;
for (i = 1; i <= n_dis; i++) to_txt << SW(3) << O[i].I << SW(4) << O[i].J << SP(11) << SW(22) << O[i].D << endl;
to_txt << endl;

to_txt << "Direction observations:" << endl;
to_txt << "Pt1 Pt2                  Direction    [deg, dms]" << endl;
to_txt << "-----------------------------------------------" << endl;
for (i = n_dis + 1; i <= n_obs; i++) to_txt << SW(3) << O[i].I << SW(4) << O[i].J << SP(13) << SW(20) << 180 * O[i].D / pi << SW(20) << dms(180 * O[i].D
    / pi) << endl;
to_txt << endl;

// ******************************************************************************************************
// Assigning weight matrix (mP)

// Distances:
for (i = 1; i <= n_dis; i++) mP(i, i) = 1 / sq(0.069);                                  // a priori sigma for distances is 0.069 meters

// Directions:
for (i = n_dis + 1; i <= n_obs; i++) mP(i, i) = 1 / sq(pi * deg(0.000011) / 180);              // a priori sigma for directions is 0.11 arcseconds

int iter;           // adjustment iteration counter

// ******************************************************************************************************
// ******************************************************************************************************

for (iter = 1; iter <= it; iter++)
{
    to_txt << iter;
    if (iter % 100 == 1) to_txt << "st";
    else if (iter % 100 == 2) to_txt << "nd";
    else if (iter % 100 == 3) to_txt << "rd";
    else to_txt << "th";
    to_txt << " iteration:" << endl << endl;

    // ------------------------------------------------------------------------------------------------------
    // Creating network design matrix (mA) and misclosure vector (mL)
```

```
if (cs == 1)    // Geodetic Coordinate System
{
    // improved geodetic coordinates ...
    for (i = 1; i <= n_new; i++) aNP[i] = { vX(2 * i - 1), vX(2 * i), NP[i].h , "" };
    for (i = 1; i <= n_obs; i++)
    {
        // current geodetic coordinates of standpoint & forepoint:
        if (O[i].I <= n_new) I = aNP[O[i].I];                                    // I is new network point
        else I = NP[O[i].I];                                                      // I is fixed network point
        if (O[i].J <= n_new) J = aNP[O[i].J];                                    // J is new network point
        else J = NP[O[i].J];                                                      // J is fixed network point
        if (i <= n_dis)        // Distances
        {
            if (O[i].I <= n_new) mA(i, 2 * O[i].I - 1) = Dis_dv_Il(I, J), mA(i, 2 * O[i].I) = Dis_dv_If(I, J);
            if (O[i].J <= n_new) mA(i, 2 * O[i].J - 1) = Dis_dv_Jl(I, J), mA(i, 2 * O[i].J) = Dis_dv_Jf(I, J);
            vL(i) = O[i].D - distance_IJ(I, J);
        }
        else                   // Directions
        {
            if (O[i].I <= n_new) mA(i, 2 * O[i].I - 1) = Dir_dv_Il(I, J), mA(i, 2 * O[i].I) = Dir_dv_If(I, J);
            if (O[i].J <= n_new) mA(i, 2 * O[i].J - 1) = Dir_dv_Jl(I, J), mA(i, 2 * O[i].J) = Dir_dv_Jf(I, J);
            mA(i, 2 * n_new + O[i].I) = -1.0;
            vL(i) = O[i].D - azimuth_IJ(I, J) + vX(2 * n_new + O[i].I);
            if (vL(i) < -pi) vL(i) += 2 * pi;
            else if (vL(i) > pi) vL(i) -= 2 * pi;
        }
    }
}
else            // Projected Coordinate System
{
    // improved geodetic coordinates ...
    for (i = 1; i <= n_new; i++)
    {
        if (cs == 2) aNP[i] = lf_from_enCC(vX(2 * i - 1), vX(2 * i), aNP[i], NC);
        if (cs == 3) aNP[i] = lf_from_enEAC(vX(2 * i - 1), vX(2 * i), aNP[i], NC);
        if (cs == 4) aNP[i] = lf_from_enTM(vX(2 * i - 1), vX(2 * i), aNP[i], L0, K0, E0, N0);
    }
    for (i = 1; i <= n_obs; i++)
    {
        // current projected and geodetic coordinates of standpoint & forepoint:
        if (O[i].I <= n_new) ei = vX(2 * O[i].I - 1), ni = vX(2 * O[i].I), I = aNP[O[i].I];  // I is new network point
        else ei = eNP[O[i].I], ni = nNP[O[i].I], I = NP[O[i].I];                              // I is fixed network point
        if (O[i].J <= n_new) ej = vX(2 * O[i].J - 1), nj = vX(2 * O[i].J), J = aNP[O[i].J];  // J is new network point
        else ej = eNP[O[i].J], nj = nNP[O[i].J], J = NP[O[i].J];                              // J is fixed network point
```

```
        if (i <= n_dis)         // Distances
        {
            if (O[i].I <= n_new) mA(i, 2 * O[i].I - 1) = Dis_dv_Ie(ei, ni, ej, nj), mA(i, 2 * O[i].I) = Dis_dv_In(ei, ni, ej, nj);
            if (O[i].J <= n_new) mA(i, 2 * O[i].J - 1) = Dis_dv_Je(ei, ni, ej, nj), mA(i, 2 * O[i].J) = Dis_dv_Jn(ei, ni, ej, nj);
            vL(i) = O[i].D - distance_IJ(I, J);
        }
        else                    // Directions
        {
            if (O[i].I <= n_new) mA(i, 2 * O[i].I - 1) = Dir_dv_Ie(ei, ni, ej, nj), mA(i, 2 * O[i].I) = Dir_dv_In(ei, ni, ej, nj);
            if (O[i].J <= n_new) mA(i, 2 * O[i].J - 1) = Dir_dv_Je(ei, ni, ej, nj), mA(i, 2 * O[i].J) = Dir_dv_Jn(ei, ni, ej, nj);
            mA(i, 2 * n_new + O[i].I) = -1.0;
            vL(i) = O[i].D - azimuth_IJ(I, J) + vX(2 * n_new + O[i].I);
            if (vL(i) < -pi) vL(i) += 2 * pi;
            else if (vL(i) > pi) vL(i) -= 2 * pi;
        }
    }
}

to_txt << "    Distance misclosures:" << endl;
to_txt << "    Pt1 Pt2  Distance MIS [m]" << endl;
to_txt << "    -------------------------" << endl;
for (i = 1; i <= n_dis; i++) to_txt << SW(7) << O[i].I << SW(4) << O[i].J << SP(11) << SW(18) << vL(i) << endl;
to_txt << endl;
to_txt << "    Direction misclosures:" << endl;
to_txt << "    Pt1 Pt2            Direction MIS       [deg, dms]" << endl;
to_txt << "    ---------------------------------------------------" << endl;
for (i = n_dis + 1; i <= n_obs; i++) to_txt << SW(7) << O[i].I << SW(4) << O[i].J << SP(16) << SW(22) << 180 * vL(i) / pi << SW(22) << dms(180 * vL(i)
    / pi) << endl;
to_txt << endl;

// -----------------------------------------------------------------------------------------------------------------------
// Vector of corrections of unknowns (vdX) and vector of estimated values of unknowns (vX)

vdX = tr(mA) * mP * vL;        // temporarily
mQ = tr(mA) * mP * mA;         // temporarily, normal equation matrix
mQ = ps(mQ);                   // cofactor matrix of unknowns (pseudoinversion is used here to achieve a bit higher accuracy)
vdX = mQ * vdX;                // vector of corrections of unknowns
vX = vX + vdX;                 // vector of estimated values of unknowns

for (i = 2 * n_new + 1; i <= n_unk; i++) if (vX(i) < 0 || vX(i) > 2 * pi) vX(i) = angle_between_0_and_2pi(vX(i));
```

```
to_txt << "    Coordinate corrections of new network points:" << endl;
if (cs == 1)    // Geodetic Coordinate System
{
    to_txt << "     Pt           Longitude COR       [deg, dms]              Latitude COR       [deg, dms] Hz COR(X,Y,Z) [m]" << endl;
    to_txt << "    ----------------------------------------------------------------------------------------------------------" << endl;
    for (i = 1; i <= n_new; i++)
    {
        I = { vX(2 * i - 1), vX(2 * i), NP[i].h , "" };
        to_txt << SW(7) << i << SP(16) << SW(21) << 180 * vdX(2 * i - 1) / pi << SW(21) << dms(180 * vdX(2 * i - 1) / pi) << SW(21) << 180 * vdX(2 * i)
            / pi << SW(21) << dms(180 * vdX(2 * i) / pi) << SP(11) << SW(18) << distance_IJ(I, aNP[i]) << endl;
    }
}
else            // Projected Coordinate System
{
    to_txt << "     Pt   Easting COR [m]  Northing COR [m]   Hz COR(e,n) [m] Hz COR(X,Y,Z) [m]" << endl;
    to_txt << "    ----------------------------------------------------------------------------" << endl;
    for (i = 1; i <= n_new; i++)
    {
        if (cs == 2) I = lf_from_enCC(vX(2 * i - 1), vX(2 * i), aNP[i], NC);
        if (cs == 3) I = lf_from_enEAC(vX(2 * i - 1), vX(2 * i), aNP[i], NC);
        if (cs == 4) I = lf_from_enTM(vX(2 * i - 1), vX(2 * i), aNP[i], L0, K0, E0, N0);
        to_txt << SW(7) << i << SP(11) << SW(18) << vdX(2 * i - 1) << SW(18) << vdX(2 * i) << SW(18) << sqrtl(sq(vdX(2 * i - 1)) + sq(vdX(2 * i)))
            << SW(18) << distance_IJ(I, aNP[i]) << endl;
    }
}
to_txt << endl;

to_txt << "    Orientation angle corrections:" << endl;
to_txt << "     Pt          Orientation COR      [deg, dms]" << endl;
to_txt << "    ----------------------------------------------" << endl;
for (i = 2 * n_new + 1; i <= n_unk; i++) to_txt << SW(6) << i - 2 * n_new << SP(16) << SW(22) << 180 * vdX(i) / pi << SW(22) << dms(180 * vdX(i) / pi)
    << endl;
to_txt << endl;

to_txt << "    Improved coordinates of new network points:" << endl;
if (cs == 1)    // Geodetic Coordinate System
{
    to_txt << "     Pt                Longitude    [deg, dms]                 Latitude    [deg, dms]" << endl;
    to_txt << "    ----------------------------------------------------------------------------------" << endl;
    for (i = 1; i <= n_new; i++) to_txt << SW(7) << i << SP(13) << SW(20) << 180 * vX(2 * i - 1) / pi << SW(20) << dms(180 * vX(2 * i - 1) / pi)
        << SW(20) << 180 * vX(2 * i) / pi << SW(20) << dms(180 * vX(2 * i) / pi) << endl;
}
```

```
    else            // Projected Coordinate System
    {
        to_txt << "     Pt          Easting [m]          Northing [m]" << endl;
        to_txt << "    ----------------------------------------------" << endl;
        for (i = 1; i <= n_new; i++) to_txt << SW(7) << i << SP(11) << SW(22) << vX(2 * i - 1) << SW(22) << vX(2 * i) << endl;
    }
    to_txt << endl;
    to_txt << "    Improved orientation angles:" << endl;
    to_txt << "     Pt              Orientation    [deg, dms]" << endl;
    to_txt << "    ------------------------------------------" << endl;
    for (i = 1; i <= n_pts; i++) to_txt << SW(7) << i << SP(13) << SW(20) << 180 * vX(2 * n_new + i) / pi << SW(20) << dms(180 * vX(2 * n_new + i) / pi)
        << endl;
    to_txt << endl;
}

// *****************************************************************************************************************************************************
// *****************************************************************************************************************************************************

to_txt << it << " iteration";
if (it > 1) to_txt << "s";
to_txt << " completed!" << endl << endl;

// -----------------------------------------------------------------------------------------------------------------------------------------------------
// Estimated values of the network unknowns

to_txt << "Final coordinates of new network points:" << endl;
if (cs == 1)        // Geodetic Coordinate System
{
    to_txt << " Pt                Longitude    [deg, dms]                 Latitude    [deg, dms]" << endl;
    to_txt << "---------------------------------------------------------------------------------" << endl;
    for (i = 1; i <= n_new; i++) to_txt << SW(3) << i << SP(13) << SW(20) << 180 * vX(2 * i - 1) / pi << SW(20) << dms(180 * vX(2 * i - 1) / pi) << SW(20)
        << 180 * vX(2 * i) / pi << SW(20) << dms(180 * vX(2 * i) / pi) << endl;
}
else                // Projected Coordinate System
{
    to_txt << " Pt          Easting [m]          Northing [m]" << endl;
    to_txt << "----------------------------------------------" << endl;
    for (i = 1; i <= n_new; i++) to_txt << SW(3) << i << SP(11) << SW(22) << vX(2 * i - 1) << SW(22) << vX(2 * i) << endl;
}
to_txt << endl;
```

```
to_txt << "Final orientation angles:" << endl;
to_txt << " Pt               Orientation    [deg, dms]" << endl;
to_txt << "-------------------------------------------" << endl;
for (i = 1; i <= n_pts; i++) to_txt << SW(3) << i << SP(13) << SW(20) << 180 * vX(2 * n_new + i) / pi << SW(20) << dms(180 * vX(2 * n_new + i) / pi)
    << endl;
to_txt << endl;

// -------------------------------------------------------------------------------------------------------------------------------------------------------
// Errors of the estimated network unknowns - comparing to their exact values

to_txt << "Coordinate errors of new network points:" << endl;
if (cs == 1)        // Geodetic Coordinate System
{
    to_txt << " Pt            Longitude ERR       [deg, dms]             Latitude ERR       [deg, dms] Hz ERR(X,Y,Z) [m]" << endl;
    to_txt << "---------------------------------------------------------------------------------------------------------" << endl;
    for (i = 1; i <= n_new; i++)
    {
        I = { vX(2 * i - 1), vX(2 * i), NP[i].h , "" };
        to_txt << SW(3) << i << SP(16) << SW(21) << 180 * (NP[i].l - I.l) / pi << SW(21) << dms(180 * (NP[i].l - I.l) / pi) << SW(21) << 180 * (NP[i].f
            - I.f) / pi << SW(21) << dms(180 * (NP[i].f - I.f) / pi) << SP(11) << SW(18) << distance_IJ(I, NP[i]) << endl;
    }
}
else                // Projected Coordinate System
{
    to_txt << " Pt   Easting ERR [m]  Northing ERR [m]   Hz ERR(e,n) [m] Hz ERR(X,Y,Z) [m]" << endl;
    to_txt << "---------------------------------------------------------------------------" << endl;
    for (i = 1; i <= n_new; i++)
    {
        to_txt << SW(3) << i << SP(11) << SW(18) << eNP[i] - vX(2 * i - 1) << SW(18) << nNP[i] - vX(2 * i) << SW(18) << grid_distance_IJ(eNP[i], nNP[i],
            vX(2 * i - 1), vX(2 * i));
        // final geodetic coordinates ...
        if (cs == 2) aNP[i] = lf_from_enCC(vX(2 * i - 1), vX(2 * i), aNP[i], NC);
        if (cs == 3) aNP[i] = lf_from_enEAC(vX(2 * i - 1), vX(2 * i), aNP[i], NC);
        if (cs == 4) aNP[i] = lf_from_enTM(vX(2 * i - 1), vX(2 * i), aNP[i], L0, K0, E0, N0);
        to_txt << SW(18) << distance_IJ(aNP[i], NP[i]) << endl;
    }
}
to_txt << endl;
```

```
to_txt << "Orientation angle errors:" << endl;
to_txt << " Pt            Orientation ERR       [deg, dms]" << endl;
to_txt << "-----------------------------------------------" << endl;
for (i = 1; i <= n_pts; i++)
{
    tmp = (OR[i] - vX(2 * n_new + i));
    if (tmp < -pi) tmp += 2 * pi;
    else if (tmp > pi) tmp -= 2 * pi;
    to_txt << SW(3) << i << SP(16) << SW(22) << 180 * tmp / pi << SW(22) << dms(180 * tmp / pi) << endl;
}
to_txt << endl;

// ------------------------------------------------------------------------------------------------------------------
// Vector of observation residuals (vV)

vV = mA * vdX - vL;

to_txt << "Distance residuals:" << endl;
to_txt << "Pt1 Pt2  Distance RES [m]" << endl;
to_txt << "-------------------------" << endl;
for (i = 1; i <= n_dis; i++) to_txt << SW(3) << O[i].I << SW(4) << O[i].J << SP(11) << SW(18) << vV(i) << endl;
to_txt << endl;

to_txt << "Direction residuals:" << endl;
to_txt << "Pt1 Pt2              Direction RES       [deg, dms]" << endl;
to_txt << "---------------------------------------------------" << endl;
for (i = n_dis + 1; i <= n_obs; i++) to_txt << SW(3) << O[i].I << SW(4) << O[i].J << SP(16) << SW(22) << 180 * vV(i) / pi << SW(22) << dms(180 * vV(i)
    / pi) << endl;
to_txt << endl;

full s2;            // estimated variance factor (a posteriori)

s2 = (tr(vV) * mP * vV)(1) / n_red;

to_txt << "Variance of unit weight = " << SP(16) << s2 << endl;
to_txt << "Standard deviation of unit weight = " << SP(16) << sqrtl(s2) << endl << endl;

// ---------------------------------------------------------------------------------------------------------------
// Variance-covariance matrix of the network unknowns (mΣ)

mQ = s2 * mQ;
```

```
to_txt << "Coordinate standard deviations of new network points:" << endl;
if (cs == 1)        // Geodetic Coordinate System
{
    to_txt << " Pt            Longitude STD       [deg, dms]             Latitude STD       [deg, dms]" << endl;
    to_txt << "-----------------------------------------------------------------------------------------" << endl;
    for (i = 1; i <= n_new; i++)
    {
        tmp = sqrtl(mQ(2 * i - 1, 2 * i - 1));
        to_txt << SW(3) << i << SP(16) << SW(21) << 180 * tmp / pi << SW(21) << dms(180 * tmp / pi);
        tmp = sqrtl(mQ(2 * i, 2 * i));
        to_txt << SW(21) << 180 * tmp / pi << SW(21) << dms(180 * tmp / pi) << endl;
    }
}
else                // Projected Coordinate System
{
    to_txt << " Pt   Easting STD [m]  Northing STD [m]  Position STD [m]" << endl;
    to_txt << "----------------------------------------------------------" << endl;
    for (i = 1; i <= n_new; i++)
    {
        to_txt << SW(3) << i << SP(11) << SW(18) << sqrtl(mQ(2 * i - 1, 2 * i - 1)) << SW(18) << sqrtl(mQ(2 * i, 2 * i)) << SW(18)
            << sqrtl(mQ(2 * i - 1, 2 * i - 1) + mQ(2 * i, 2 * i)) << endl;
    }
}
to_txt << endl;

to_txt << "Orientation standard deviations:" << endl;
to_txt << " Pt            Orientation STD       [deg, dms]" << endl;
to_txt << "------------------------------------------------" << endl;
for (i = 1; i <= n_pts; i++)
{
    tmp = sqrtl(mQ(2 * n_new + i, 2 * n_new + i));
    to_txt << SW(3) << i << SP(16) << SW(22) << 180 * tmp / pi << SW(22) << dms(180 * tmp / pi) << endl;
}
to_txt << endl;
```

```
confidence_ellipse ellipse;
if (cs == 1)
{
    // Creating metric or Lamé matrix (mH) (Soler & Smith 2010), but referring to the ellipsoid (h = 0)
    for (i = 1; i <= n_new; i++)
    {
        mH(2 * i - 1, 2 * i - 1) = Nc(vX(2 * i)) * cosl(vX(2 * i));
        mH(2 * i, 2 * i) = Mc(vX(2 * i));
    }
    for (i = 2 * n_new + 1; i <= n_unk; i++) mH(i, i) = 1.0;
    mQl = mH * mQ * mH;              // metric variance-covariance matrix of unknowns in local geodetic systems (for each new network point)
}

to_txt << "Standard confidence ellipse elements of new network points";
if (cs == 1) to_txt << endl << "(local geodetic system (e, n) for each individual network point)";
to_txt << ":" << endl;
to_txt << " Pt  Major S-Axis [m]  Minor S-Axis [m]   Major S-Axis Azimuth [deg, dms]" << endl;
to_txt << "--------------------------------------------------------------------------" << endl;
for (i = 1; i <= n_new; i++)
{
    if (cs == 1) ellipse = confidence_ellipse(mQl(2 * i - 1, 2 * i - 1), mQl(2 * i, 2 * i), mQl(2 * i - 1, 2 * i));
    else ellipse = confidence_ellipse(mQ(2 * i - 1, 2 * i - 1), mQ(2 * i, 2 * i), mQ(2 * i - 1, 2 * i));
    to_txt << SW(3) << i << SP(11) << SW(18) << ellipse.a << SW(18) << ellipse.b << SP(10) << SW(17) << 180 * ellipse.t / pi << SW(17) << dms(180
        * ellipse.t / pi) << endl;
}
to_txt << endl;

if (cs == 1)
{
    matrix mAp(n_obs, n_unk);        // network design matrix for the computation in the projected coordinate system
    matrix vOp(n_obs);               // vector of reduced observations for the computation in the projected coordinate system
    matrix vVp(n_obs);               // vector of observation residuals for the computation in the projected coordinate system

    to_txt << "CONVERSION of the network adjustment results into the projected coordinate systems" << endl << endl;
    to_txt << "1 ... Conformal Cylindrical projection" << endl;
    to_txt << "---------------------------------------" << endl << endl;

    to_txt << "Final coordinates of new network points:" << endl;
    to_txt << " Pt           Easting [m]          Northing [m]" << endl;
    to_txt << "------------------------------------------------" << endl;
    for (i = 1; i <= n_new; i++)
    {
        I = { vX(2 * i - 1), vX(2 * i), aNP[i].h, "" };
        to_txt << SW(3) << i << SP(11) << SW(22) << eCC(I, NC) << SW(22) << nCC(I, NC) << endl;
    }
```

```
to_txt << endl;
// Creating network design matrix (mAp), vector of reduced observations (vOp), and vector of observation reductions (vR)
// for the computation in the projected coordinate system:
for (i = 1; i <= n_obs; i++)
{
    // final projected and geodetic coordinates of standpoint & forepoint:
    if (O[i].I <= n_new) I = aNP[O[i].I], ei = eCC(I, NC), ni = nCC(I, NC);      // I is new network point
    else I = NP[O[i].I], ei = eCC(I, NC), ni = nCC(I, NC);                        // I is fixed network point
    if (O[i].J <= n_new) J = aNP[O[i].J], ej = eCC(J, NC), nj = nCC(J, NC);      // J is new network point
    else J = NP[O[i].J], ej = eCC(J, NC), nj = nCC(J, NC);                        // J is fixed network point
    if (i <= n_dis)         // Distances
    {
        if (O[i].I <= n_new) mAp(i, 2 * O[i].I - 1) = Dis_dv_Ie(ei, ni, ej, nj), mAp(i, 2 * O[i].I) = Dis_dv_In(ei, ni, ej, nj);
        if (O[i].J <= n_new) mAp(i, 2 * O[i].J - 1) = Dis_dv_Je(ei, ni, ej, nj), mAp(i, 2 * O[i].J) = Dis_dv_Jn(ei, ni, ej, nj);
        vR(i) = grid_distance_IJ(ei, ni, ej, nj) - distance_IJ(I, J);
        vOp(i) = O[i].D + vR(i);
    }
    else                    // Directions
    {
        if (O[i].I <= n_new) mAp(i, 2 * O[i].I - 1) = Dir_dv_Ie(ei, ni, ej, nj), mAp(i, 2 * O[i].I) = Dir_dv_In(ei, ni, ej, nj);
        if (O[i].J <= n_new) mAp(i, 2 * O[i].J - 1) = Dir_dv_Je(ei, ni, ej, nj), mAp(i, 2 * O[i].J) = Dir_dv_Jn(ei, ni, ej, nj);
        mAp(i, 2 * n_new + O[i].I) = -1.0;
        vR(i) = grid_azimuth_IJ(ei, ni, ej, nj) - azimuth_IJ(I, J);
        vOp(i) = O[i].D + vR(i);
    }
}

to_txt << "Reduced distance observations:" << endl;
to_txt << "Pt1 Pt2  Reduced distance [m]" << endl;
to_txt << "-----------------------------" << endl;
for (i = 1; i <= n_dis; i++) to_txt << SW(3) << O[i].I << SW(4) << O[i].J << SP(11) << SW(22) << vOp(i) << endl;
to_txt << endl;

to_txt << "Reduced direction observations:" << endl;
to_txt << "Pt1 Pt2         Reduced direction    [deg, dms]" << endl;
to_txt << "-----------------------------------------------" << endl;
for (i = n_dis + 1; i <= n_obs; i++) to_txt << SW(3) << O[i].I << SW(4) << O[i].J << SP(13) << SW(20) << 180 * vOp(i) / pi << SW(20)
    << dms(180 * vOp(i) / pi) << endl;
to_txt << endl;
```

```
to_txt << "Grid azimuths:" << endl;
to_txt << "Pt1 Pt2             Grid azimuth    [deg, dms]" << endl;
to_txt << "-----------------------------------------------" << endl;
for (i = n_dis + 1; i <= n_obs; i++)
{
    if (O[i].I <= n_new) I = aNP[O[i].I], ei = eCC(I, NC), ni = nCC(I, NC);      // I is new network point
    else I = NP[O[i].I], ei = eCC(I, NC), ni = nCC(I, NC);                       // I is fixed network point
    if (O[i].J <= n_new) J = aNP[O[i].J], ej = eCC(J, NC), nj = nCC(J, NC);      // J is new network point
    else J = NP[O[i].J], ej = eCC(J, NC), nj = nCC(J, NC);                       // J is fixed network point
    tmp = grid_azimuth_IJ(ei, ni, ej, nj);
    to_txt << SW(3) << O[i].I << SW(4) << O[i].J << SP(13) << SW(20) << 180 * tmp / pi << SW(20) << dms(180 * tmp / pi) << endl;
}
to_txt << endl;

to_txt << "Distance reductions:" << endl;
to_txt << "Pt1 Pt2  Distance RED [m]" << endl;
to_txt << "-------------------------" << endl;
for (i = 1; i <= n_dis; i++) to_txt << SW(3) << O[i].I << SW(4) << O[i].J << SP(11) << SW(18) << vR(i) << endl;
to_txt << endl;

to_txt << "Direction reductions:" << endl;
to_txt << "Pt1 Pt2             Direction RED       [deg, dms]" << endl;
to_txt << "---------------------------------------------------" << endl;
for (i = n_dis + 1; i <= n_obs; i++) to_txt << SW(3) << O[i].I << SW(4) << O[i].J << SP(16) << SW(22) << 180 * vR(i) / pi << SW(22)
    << dms(180 * vR(i) / pi) << endl;
to_txt << endl;

// Creating mapping design matrix (mAm):
for (i = 1; i <= n_new; i++)
{
    I = { vX(2 * i - 1), vX(2 * i), aNP[i].h, "" };
    mAm(2 * i - 1, 2 * i - 1) = eCC_dv_l(I, NC), mAm(2 * i - 1, 2 * i) = 0.0;
    mAm(2 * i, 2 * i - 1) = 0.0, mAm(2 * i, 2 * i) = nCC_dv_f(I, NC);
}
for (i = 2 * n_new + 1; i <= n_unk; i++) mAm(i, i) = 1.0;
mQp = mAm * mQ * tr(mAm);           // variance-covariance matrix of unknowns in the projected coordinate system – CC projection
```

```
to_txt << "Standard confidence ellipse elements of new network points (I):" << endl;
to_txt << " Pt  Major S-Axis [m]  Minor S-Axis [m]   Major S-Axis Azimuth [deg, dms]" << endl;
to_txt << "-------------------------------------------------------------------------" << endl;
for (i = 1; i <= n_new; i++)
{
    ellipse = confidence_ellipse(mQp(2 * i - 1, 2 * i - 1), mQp(2 * i, 2 * i), mQp(2 * i - 1, 2 * i));
    to_txt << SW(3) << i << SP(11) << SW(18) << ellipse.a << SW(18) << ellipse.b << SP(10) << SW(17) << 180 * ellipse.t / pi
       << SW(17) << dms(180 * ellipse.t / pi) << endl;
}
to_txt << endl;

// ... the same as above, but using scale factor and meridian convergence instead of Jacobian matrices ...
to_txt << "Standard confidence ellipse elements of new network points (II):" << endl;
to_txt << " Pt  Major S-Axis [m]  Minor S-Axis [m]   Major S-Axis Azimuth [deg, dms]" << endl;
to_txt << "-------------------------------------------------------------------------" << endl;
for (i = 1; i <= n_new; i++)
{
    I = { vX(2 * i - 1), vX(2 * i), 0.0 , "" };
    ellipse = confidence_ellipse(mQl(2 * i - 1, 2 * i - 1), mQl(2 * i, 2 * i), mQl(2 * i - 1, 2 * i));
    tmp = CC_scale(I, NC);     // scale factor is used to adapt the confidence ellipse estimated in the local geodetic system
    to_txt << SW(3) << i << SP(11) << SW(18) << ellipse.a * tmp << SW(18) << ellipse.b * tmp;
    tmp = 0.0;                 // meridian convergence is used to rotate the confidence ellipse estimated in the local geodetic system
    if (ellipse.a == ellipse.b) to_txt << SP(10) << SW(17) << 0.0 << SW(17) << 0.0 << endl;
    else to_txt << SP(10) << SW(17) << 180 * (ellipse.t - tmp) / pi << SW(17) << dms(180 * (ellipse.t - tmp) / pi) << endl;
}
to_txt << endl;

to_txt << "Scale factors and distortions at network points:" << endl;
to_txt << " Pt       Scale factor []      Distortion [‰]" << endl;
to_txt << "---------------------------------------------" << endl;
for (i = 1; i <= n_pts; i++)
{
    I = NP[i];
    to_txt << SW(3) << i << SP(16) << SW(22) << CC_scale(I, NC) << SP (13) << SW(20) << 1000.0 * (1.0 - CC_scale(I, NC)) << endl;
}
to_txt << endl;
```

```
to_txt << "2 ... Equal-Area Cylindrical projection" << endl;
to_txt << "---------------------------------------" << endl << endl;

to_txt << "Final coordinates of new network points:" << endl;
to_txt << " Pt             Easting [m]           Northing [m]" << endl;
to_txt << "-------------------------------------------------" << endl;
for (i = 1; i <= n_new; i++)
{
    I = { vX(2 * i - 1), vX(2 * i), aNP[i].h, "" };
    to_txt << SW(3) << i << SP(11) << SW(22) << eEAC(I, NC) << SW(22) << nEAC(I, NC) << endl;
}
to_txt << endl;

// Creating network design matrix (mAp), vector of reduced observations (vOp), and vector of observation reductions (vR)
// for the computation in the projected coordinate system:
for (i = 1; i <= n_obs; i++)
{
    // final projected and geodetic coordinates of standpoint & forepoint:
    if (O[i].I <= n_new) I = aNP[O[i].I], ei = eEAC(I, NC), ni = nEAC(I, NC);    // I is new network point
    else I = NP[O[i].I], ei = eEAC(I, NC), ni = nEAC(I, NC);                     // I is fixed network point
    if (O[i].J <= n_new) J = aNP[O[i].J], ej = eEAC(J, NC), nj = nEAC(J, NC);    // J is new network point
    else J = NP[O[i].J], ej = eEAC(J, NC), nj = nEAC(J, NC);                     // J is fixed network point
    if (i <= n_dis)        // Distances
    {
        if (O[i].I <= n_new) mAp(i, 2 * O[i].I - 1) = Dis_dv_Ie(ei, ni, ej, nj), mAp(i, 2 * O[i].I) = Dis_dv_In(ei, ni, ej, nj);
        if (O[i].J <= n_new) mAp(i, 2 * O[i].J - 1) = Dis_dv_Je(ei, ni, ej, nj), mAp(i, 2 * O[i].J) = Dis_dv_Jn(ei, ni, ej, nj);
        vR(i) = grid_distance_IJ(ei, ni, ej, nj) - distance_IJ(I, J);
        vOp(i) = O[i].D + vR(i);
    }
    else                   // Directions
    {
        if (O[i].I <= n_new) mAp(i, 2 * O[i].I - 1) = Dir_dv_Ie(ei, ni, ej, nj), mAp(i, 2 * O[i].I) = Dir_dv_In(ei, ni, ej, nj);
        if (O[i].J <= n_new) mAp(i, 2 * O[i].J - 1) = Dir_dv_Je(ei, ni, ej, nj), mAp(i, 2 * O[i].J) = Dir_dv_Jn(ei, ni, ej, nj);
        mAp(i, 2 * n_new + O[i].I) = -1.0;
        vR(i) = grid_azimuth_IJ(ei, ni, ej, nj) - azimuth_IJ(I, J);
        vOp(i) = O[i].D + vR(i);
    }
}

to_txt << "Reduced distance observations:" << endl;
to_txt << "Pt1 Pt2  Reduced distance [m]" << endl;
to_txt << "-----------------------------" << endl;
for (i = 1; i <= n_dis; i++) to_txt << SW(3) << O[i].I << SW(4) << O[i].J << SP(11) << SW(22) << vOp(i) << endl;
to_txt << endl;
```

```
to_txt << "Reduced direction observations:" << endl;
to_txt << "Pt1 Pt2         Reduced direction    [deg, dms]" << endl;
to_txt << "-----------------------------------------------" << endl;
for (i = n_dis + 1; i <= n_obs; i++) to_txt << SW(3) << O[i].I << SW(4) << O[i].J << SP(13) << SW(20) << 180 * vOp(i) / pi << SW(20)
    << dms(180 * vOp(i) / pi) << endl;
to_txt << endl;

to_txt << "Grid azimuths:" << endl;
to_txt << "Pt1 Pt2              Grid azimuth    [deg, dms]" << endl;
to_txt << "-----------------------------------------------" << endl;
for (i = n_dis + 1; i <= n_obs; i++)
{
    if (O[i].I <= n_new) I = aNP[O[i].I], ei = eEAC(I, NC), ni = nEAC(I, NC);    // I is new network point
    else I = NP[O[i].I], ei = eEAC(I, NC), ni = nEAC(I, NC);                     // I is fixed network point
    if (O[i].J <= n_new) J = aNP[O[i].J], ej = eEAC(J, NC), nj = nEAC(J, NC);    // J is new network point
    else J = NP[O[i].J], ej = eEAC(J, NC), nj = nEAC(J, NC);                     // J is fixed network point
    tmp = grid_azimuth_IJ(ei, ni, ej, nj);
    to_txt << SW(3) << O[i].I << SW(4) << O[i].J << SP(13) << SW(20) << 180 * tmp / pi << SW(20) << dms(180 * tmp / pi) << endl;
}
to_txt << endl;

to_txt << "Distance reductions:" << endl;
to_txt << "Pt1 Pt2  Distance RED [m]" << endl;
to_txt << "-------------------------" << endl;
for (i = 1; i <= n_dis; i++) to_txt << SW(3) << O[i].I << SW(4) << O[i].J << SP(11) << SW(18) << vR(i) << endl;
to_txt << endl;

to_txt << "Direction reductions:" << endl;
to_txt << "Pt1 Pt2              Direction RED       [deg, dms]" << endl;
to_txt << "---------------------------------------------------" << endl;
for (i = n_dis + 1; i <= n_obs; i++) to_txt << SW(3) << O[i].I << SW(4) << O[i].J << SP(16) << SW(22) << 180 * vR(i) / pi << SW(22)
    << dms(180 * vR(i) / pi) << endl;
to_txt << endl;

// Creating mapping design matrix (mAm):
for (i = 1; i <= n_new; i++)
{
    I = { vX(2 * i - 1), vX(2 * i), aNP[i].h, "" };
    mAm(2 * i - 1, 2 * i - 1) = eEAC_dv_l(I, NC), mAm(2 * i - 1, 2 * i) = 0.0;
    mAm(2 * i, 2 * i - 1) = 0.0, mAm(2 * i, 2 * i) = nEAC_dv_f(I, NC);
}
for (i = 2 * n_new + 1; i <= n_unk; i++) mAm(i, i) = 1.0;
mQp = mAm * mQ * tr(mAm);           // variance-covariance matrix of unknowns in the projected coordinate system - EAC projection
```

```
to_txt << "Standard confidence ellipse elements of new network points:" << endl;
to_txt << " Pt  Major S-Axis [m]  Minor S-Axis [m]   Major S-Axis Azimuth [deg, dms]" << endl;
to_txt << "-------------------------------------------------------------------------" << endl;
for (i = 1; i <= n_new; i++)
{
    ellipse = confidence_ellipse(mQp(2 * i - 1, 2 * i - 1), mQp(2 * i, 2 * i), mQp(2 * i - 1, 2 * i));
    to_txt << SW(3) << i << SP(11) << SW(18) << ellipse.a << SW(18) << ellipse.b << SP(10) << SW(17) << 180 * ellipse.t / pi
       << SW(17) << dms(180 * ellipse.t / pi) << endl;
}
to_txt << endl;

to_txt << "Minimum and maximum scale factors and distortions at network points:" << endl;
to_txt << " Pt   Min scale factor []  Max distortion [‰]   Max scale factor []  Min distortion [‰]" << endl;
to_txt << "---------------------------------------------------------------------------------------" << endl;
for (i = 1; i <= n_pts; i++)
{
    I = NP[i];
    to_txt << SW(3) << i << SP(16) << SW(22) << min_EAC_scale(I, NC) << SP(13) << SW(20) << 1000.0 * (1.0 - min_EAC_scale(I, NC))
        << SP(16) << SW(22) << max_EAC_scale(I, NC) << SP(13) << SW(20) << 1000.0 * (1.0 - max_EAC_scale(I, NC)) << endl;
}
to_txt << endl;

to_txt << "3 ... Transverse Mercator projection" << endl;
to_txt << "------------------------------------" << endl << endl;

to_txt << "Final coordinates of new network points:" << endl;
to_txt << " Pt           Easting [m]          Northing [m]" << endl;
to_txt << "-----------------------------------------------" << endl;
for (i = 1; i <= n_new; i++)
{
    I = { vX(2 * i - 1), vX(2 * i), aNP[i].h, "" };
    to_txt << SW(3) << i << SP(11) << SW(22) << eTM(I, L0, K0, E0) << SW(22) << nTM(I, L0, K0, N0) << endl;
}
to_txt << endl;

// Creating network design matrix (mAp), vector of reduced observations (vOp), and vector of observation reductions (vR)
// for the computation in the projected coordinate system:
for (i = 1; i <= n_obs; i++)
{
    // final projected and geodetic coordinates of standpoint & forepoint:
    if (O[i].I <= n_new) I = aNP[O[i].I], ei = eTM(I, L0, K0, E0), ni = nTM(I, L0, K0, N0); // I is new network point
    else I = NP[O[i].I], ei = eTM(I, L0, K0, E0), ni = nTM(I, L0, K0, N0);                  // I is fixed network point
    if (O[i].J <= n_new) J = aNP[O[i].J], ej = eTM(J, L0, K0, E0), nj = nTM(J, L0, K0, N0); // J is new network point
    else J = NP[O[i].J], ej = eTM(J, L0, K0, E0), nj = nTM(J, L0, K0, N0);                  // J is fixed network point
```

```
    if (i <= n_dis)         // Distances
    {
        if (O[i].I <= n_new) mAp(i, 2 * O[i].I - 1) = Dis_dv_Ie(ei, ni, ej, nj), mAp(i, 2 * O[i].I) = Dis_dv_In(ei, ni, ej, nj);
        if (O[i].J <= n_new) mAp(i, 2 * O[i].J - 1) = Dis_dv_Je(ei, ni, ej, nj), mAp(i, 2 * O[i].J) = Dis_dv_Jn(ei, ni, ej, nj);
        vR(i) = grid_distance_IJ(ei, ni, ej, nj) - distance_IJ(I, J);
        vOp(i) = O[i].D + vR(i);
    }
    else                    // Directions
    {
        if (O[i].I <= n_new) mAp(i, 2 * O[i].I - 1) = Dir_dv_Ie(ei, ni, ej, nj), mAp(i, 2 * O[i].I) = Dir_dv_In(ei, ni, ej, nj);
        if (O[i].J <= n_new) mAp(i, 2 * O[i].J - 1) = Dir_dv_Je(ei, ni, ej, nj), mAp(i, 2 * O[i].J) = Dir_dv_Jn(ei, ni, ej, nj);
        mAp(i, 2 * n_new + O[i].I) = -1.0;
        vR(i) = grid_azimuth_IJ(ei, ni, ej, nj) - azimuth_IJ(I, J);
        vOp(i) = O[i].D + vR(i);
    }
}

to_txt << "Reduced distance observations:" << endl;
to_txt << "Pt1 Pt2  Reduced distance [m]" << endl;
to_txt << "-----------------------------" << endl;
for (i = 1; i <= n_dis; i++) to_txt << SW(3) << O[i].I << SW(4) << O[i].J << SP(11) << SW(22) << vOp(i) << endl;
to_txt << endl;

to_txt << "Reduced direction observations:" << endl;
to_txt << "Pt1 Pt2         Reduced direction    [deg, dms]" << endl;
to_txt << "-----------------------------------------------" << endl;
for (i = n_dis + 1; i <= n_obs; i++) to_txt << SW(3) << O[i].I << SW(4) << O[i].J << SP(13) << SW(20) << 180 * vOp(i) / pi << SW(20)
    << dms(180 * vOp(i) / pi) << endl;
to_txt << endl;

to_txt << "Grid azimuths:" << endl;
to_txt << "Pt1 Pt2              Grid azimuth    [deg, dms]" << endl;
to_txt << "-----------------------------------------------" << endl;
for (i = n_dis + 1; i <= n_obs; i++)
{
    if (O[i].I <= n_new) I = aNP[O[i].I], ei = eTM(I, L0, K0, E0), ni = nTM(I, L0, K0, N0); // I is new network point
    else I = NP[O[i].I], ei = eTM(I, L0, K0, E0), ni = nTM(I, L0, K0, N0);                  // I is fixed network point
    if (O[i].J <= n_new) J = aNP[O[i].J], ej = eTM(J, L0, K0, E0), nj = nTM(J, L0, K0, N0); // J is new network point
    else J = NP[O[i].J], ej = eTM(J, L0, K0, E0), nj = nTM(J, L0, K0, N0);                  // J is fixed network point
    tmp = grid_azimuth_IJ(ei, ni, ej, nj);
    to_txt << SW(3) << O[i].I << SW(4) << O[i].J << SP(13) << SW(20) << 180 * tmp / pi << SW(20) << dms(180 * tmp / pi) << endl;
}
to_txt << endl;
```

```
to_txt << "Distance reductions:" << endl;
to_txt << "Pt1 Pt2  Distance RED [m]" << endl;
to_txt << "-------------------------" << endl;
for (i = 1; i <= n_dis; i++) to_txt << SW(3) << O[i].I << SW(4) << O[i].J << SP(11) << SW(18) << vR(i) << endl;
to_txt << endl;

to_txt << "Direction reductions:" << endl;
to_txt << "Pt1 Pt2              Direction RED       [deg, dms]" << endl;
to_txt << "---------------------------------------------------" << endl;
for (i = n_dis + 1; i <= n_obs; i++) to_txt << SW(3) << O[i].I << SW(4) << O[i].J << SP(16) << SW(22) << 180 * vR(i) / pi << SW(22)
    << dms(180 * vR(i) / pi) << endl;
to_txt << endl;

to_txt << "Meridian convergences:" << endl;
to_txt << " Pt       Meridian convergence       [deg, dms]" << endl;
to_txt << "-----------------------------------------------" << endl;
for (i = 1; i <= n_pts; i++)
{
    I = NP[i];
    tmp = TM_converg(I, L0);        // meridian convergence
    to_txt << SW(3) << i << SP(16) << SW(22) << 180 * tmp / pi << SW(22) << dms(180 * tmp / pi) << endl;
}
to_txt << endl;
```

```
        to_txt << "Standard confidence ellipse elements of new network points:" << endl;
        to_txt << " Pt  Major S-Axis [m]  Minor S-Axis [m]   Major S-Axis Azimuth [deg, dms]" << endl;
        to_txt << "-------------------------------------------------------------------------" << endl;
        for (i = 1; i <= n_new; i++)
        {
            I = { vX(2 * i - 1), vX(2 * i), 0.0 , "" };
            ellipse = confidence_ellipse(mQl(2 * i - 1, 2 * i - 1), mQl(2 * i, 2 * i), mQl(2 * i - 1, 2 * i));
            tmp = TM_scale(I, L0, K0);  // scale factor is used to adapt the confidence ellipse estimated in the local geodetic system
            to_txt << SW(3) << i << SP(11) << SW(18) << ellipse.a * tmp << SW(18) << ellipse.b * tmp;
            tmp = TM_converg(I, L0);    // meridian convergence is used to rotate the confidence ellipse estimated in the local geodetic system
            if(ellipse.a == ellipse.b) to_txt << SP(10) << SW(17) << 0.0 << SW(17) << 0.0 << endl;
            else to_txt << SP(10) << SW(17) << 180 * (ellipse.t - tmp) / pi << SW(17) << dms(180 * (ellipse.t - tmp) / pi) << endl;
        }
        to_txt << endl;

        to_txt << "Scale factors and distortions at network points:" << endl;
        to_txt << " Pt        Scale factor []      Distortion [‰]" << endl;
        to_txt << "---------------------------------------------" << endl;
        for (i = 1; i <= n_pts; i++)
        {
            I = NP[i];
            to_txt << SW(3) << i << SP(16) << SW(22) << TM_scale(I, L0, K0) << SP(13) << SW(20) << 1000.0 * (1.0 - TM_scale(I, L0, K0))
                << endl;
        }
        to_txt << endl;
    }
    to_txt.close();

    return 1;
}
```

```
// MTRX.hpp (defines matrix arithmetics)
// *****************************************************************************************************************************************
// Algorithms are taken from Press et al. (1992) – Numerical recipes in C: the art of scientific computing

typedef long double full;

inline full sq(full x)
// square function
{
    return x * x;
}

inline full ab(full x)
// absolute value function
{
    return x < 0 ? -x : x;
}

inline int sgn(full x)
// signum function
{
    return (x > 0) - (x < 0);
}

class matrix
{
     private:
          full **p;                                  // pointer to matrix elements
          int v;                                     // number of matrix rows
          int s;                                     // number of matrix columns
     public:
          matrix(int=1,int=1);                       // default constructor of a matrix
          matrix(const matrix&);                     // copy constructor of a matrix
          ~matrix(void);                             // destructor of a matrix
          full& operator()(int,int=1);               // accesing a matrix element
          matrix& operator=(const matrix&);          // assigning a matrix
          matrix operator+(matrix);                  // adding a matrix
          matrix operator-(matrix);                  // subtracting a matrix
          matrix operator*(matrix);                  // matrix multiplication
          friend matrix operator*(full,matrix&);     // multiplying a matrix with a scalar
          friend matrix tr(matrix);                  // transpose of a matrix
          friend matrix in(matrix);                  // inverse of a matrix
          friend matrix ps(matrix);                  // pseudoinverse of a matrix
          friend full det(matrix&);                  // determinant of a matrix
};
```